\documentclass[a4paper,11pt]{article}
\pdfoutput=1 % if submitting a pdflatex (i.e. if having images in pdf, png or jpg format)
\usepackage{jcappub}

\usepackage{xcolor}
\usepackage{graphicx}
\graphicspath{{figln/}{fig/}}
\usepackage{amsmath,amssymb,amsfonts,mathrsfs,bm}
\usepackage[colorlinks=true,linkcolor=blue,citecolor=blue]{hyperref}

\usepackage{iftex}
\ifpdftex% Commands and packages specific to pdfLaTeX
    \usepackage{CJKutf8}
    \newcommand{\Ckai}[1]{\begin{CJK*}{UTF8}{gkai}#1\end{CJK*}}
\fi
\ifxetex% Commands and packages specific to XeLaTeX
    \usepackage{xeCJK}
    \newcommand{\Ckai}[1]{\textit{#1}}
\fi

\newcommand{\dif}{\mathrm{d}}
\newcommand{\ee}{\mathrm{e}}
\newcommand{\ii}{\mathrm{i}}

\newcommand{\qn}{n}
\newcommand{\ql}{\ell}
\newcommand{\qm}{m}

\newcommand{\Msun}{\,M_{\odot}}

\newcommand{\eV}{\,\mathrm{eV}}

\newcommand{\pc}{\,\mathrm{pc}}
\newcommand{\kpc}{\,\mathrm{kpc}}

\newcommand{\PBH}{\mathrm{PBH}}
\newcommand{\NFW}{\mathrm{NFW}}
\newcommand{\sol}{\mathrm{sol}}
\newcommand{\MPBH}{M_{\PBH}}
\newcommand{\NPBH}{N_{\PBH}}
\newcommand{\fPBH}{f_{\PBH}}

\begin{document}

\title{%
    Ultralight dark matter mixed with primordial black holes
}

\author[a]{Xing-Yu Yang (\Ckai{杨星宇})}
\affiliation[a]{Quantum Universe Center (QUC), Korea Institute for Advanced Study, Seoul 02455, Republic of Korea}
\emailAdd{xingyuyang@kias.re.kr}

\abstract{%
Dark matter candidates span many orders of magnitude in mass, from ultralight bosonic fields to massive compact objects.
In this work, we connect these two extremes by investigating ultralight dark matter (ULDM) mixed with primordial black holes (PBHs).
We study mixed ULDM--PBH halos by separating the continuum PBH contribution from the shot-noise fluctuation generated by discrete PBHs.
The continuum contribution enters the averaged Schr\"odinger-Poisson background, while the discreteness contribution is treated as a perturbation that induces ULDM eigenmode transitions and soliton heating.
The two contributions have distinct parametric dependencies: continuum effects scale with PBH fraction, whereas discreteness-driven transition rates scale with the product of PBH fraction and individual PBH mass in the perturbative regime.
For a fiducial mixed halo with ULDM particle mass $10^{-22}\eV$, virial mass of order $10^{10}\Msun$, and PBH fraction $1\%$, the continuum PBH component modifies the background density, gravitational potential, and low-lying ULDM eigenvalues only at the sub-percent level.
Nevertheless, this percent-level continuum PBH contribution produces a tens-of-percent response in the coherent soliton region, changing the radial mode participation by about $20\%$.
For stellar-mass PBHs, the discrete shot-noise fluctuation induces extremely slow ULDM mode transitions, with the fastest low-lying multiplet transition having a timescale of order $10^9\,\mathrm{Gyr}$ for solar-mass PBHs.
In this regime, the leading PBH effect is the continuum contribution, while discrete PBH shot noise is dynamically negligible on galactic timescales.
}

\maketitle

\section{Introduction}

Dark matter is one of the central open problems in cosmology and fundamental physics.
Although the evidence for dark matter is extensive, decades of theoretical and experimental effort have not identified its nature~\cite{Zwicky:1937zza,Rubin:1970zza,Bertone:2004pz,Clowe:2006eq,Feng:2010gw}.
Moreover, current observations constrain the total dark matter, but do not require dark matter to consist of a single component~\cite{Planck:2018vyg}.
It is therefore well motivated to consider multi-component scenarios, in which distinct dark matter candidates coexist and can leave complementary signatures~\cite{Zurek:2008qg,Marsh:2013ywa,Hertzberg:2020hsz,Schwabe:2020eac,Cai:2020fnq,Cai:2022kbp,Kim:2023onk,Yin:2024xov}.
Recent survey results also indicate sustained interest in such hybrid possibilities~\cite{Afshordi:2026arr}.

Dark matter candidates span many orders of magnitude in mass, from ultralight bosonic fields with particle masses as low as $10^{-22}\eV$ to massive compact objects as heavy as $\Msun$ or more~\cite{Cirelli:2024ssz}.
At the light end, ultralight dark matter (ULDM) candidates such as axions and axion-like particles can behave as coherent waves on galactic scales; when self-interactions are negligible, this wave-dominated limit is often referred to as fuzzy dark matter~\cite{Preskill:1982cy,Abbott:1982af,Hu:2000ke,Hui:2016ltb,Davoudiasl:2019nlo,Ferreira:2020fam,Hui:2021tkt}.
This wave nature supports solitonic cores and interference-driven granularity, giving ULDM halo structure that differs qualitatively from that of cold dark matter (CDM)~\cite{Schive:2014dra}.
At the heavy end, primordial black holes (PBHs), formed from early-universe density fluctuations, provide a compact-object dark matter candidate~\cite{Hawking:1971ei,Carr:1974nx,Khlopov:1985fch,Khlopov:2008qy,Carr:2020gox,Green:2020jor}.
PBHs have received renewed attention after the detection of gravitational waves from binary black hole mergers~\cite{Dong:2015yjs,LIGOScientific:2016aoc,Bird:2016dcv,Clesse:2016vqa,Sasaki:2016jop,Carr:2016drx,Ali-Haimoud:2016mbv,Niikura:2017zjd,Ali-Haimoud:2017rtz,Sasaki:2018dmp,Belotsky:2018wph,Cai:2019elf,Chen:2019xse,Cai:2019bmk,Carr:2020xqk,DeLuca:2020agl,Kohri:2020qqd,Kimura:2021sqz,Yuan:2021qgz,Liu:2021jnw,Wang:2022nml,Escriva:2022duf,Ozsoy:2023ryl,LISACosmologyWorkingGroup:2023njw}.

The extreme contrast between wave-like ULDM and compact-object PBHs makes their coexistence an attractive possibility for multi-component dark matter.
In this work, we study this hybrid scenario by considering dark matter halos composed of a dominant ULDM component and a subdominant PBH component.
In the limit of vanishing individual PBH mass at fixed PBH density, the PBH component is described by a continuum density, analogous to the CDM component in mixed ULDM--CDM models.
For nonzero individual PBH mass, however, the same background density profile is represented by a finite number of discrete compact objects, producing shot-noise fluctuations around that background profile.
These fluctuations can perturb the ULDM soliton and structure, even when the PBH component is subdominant.
We organize the analysis around the separation between the continuum PBH contribution and discrete PBH shot noise.
This approach delineates the parameter regimes in which a mixed ULDM--PBH halo can be modeled with a continuum treatment of the PBH component, and those in which discrete PBH shot noise can become dynamically relevant.

The paper is structured as follows.
In Section~\ref{sec:basic-framework}, we introduce the basic framework for coupled ULDM--PBH dynamics.
The continuum PBH limit and the corresponding ULDM eigenmode description are then developed in Section~\ref{sec:continuum-pbh-limit}.
We analyze discrete PBH shot noise and its resulting effects in Section~\ref{sec:discrete-pbh-effects}.
Section~\ref{sec:conclusions} summarizes the main results and discussion.

\section{Basic framework}
\label{sec:basic-framework}

We begin by formulating the coupled dynamics of a ULDM field and a PBH population.
The ULDM component is described by a classical complex field $\psi(t,\bm{x})$ that obeys the Schr\"odinger-Poisson equations~\cite{Widrow:1993qq,Schive:2014dra},
\begin{equation}
    \ii\hbar\frac{\partial\psi}{\partial t}
    =
    \left(
    -\frac{\hbar^2}{2m_\psi}\nabla^2
    +m_\psi\Phi
    \right)\psi \ ,
\end{equation}
\begin{equation}
    \nabla^2\Phi
    =
    4\pi G\left[\rho_\psi(t,\bm{x})+\rho_{\PBH}(t,\bm{x})\right] \ ,
\end{equation}
where
\begin{equation}
    \rho_\psi(t,\bm{x}) = m_\psi|\psi(t,\bm{x})|^2 \ ,
\end{equation}
and the discrete PBH density is
\begin{equation}
    \rho_{\PBH}(t,\bm{x})
    =
    \sum_{i=1}^{\NPBH}
    \MPBH \delta^{(3)}\!\left[\bm{x}-\bm{x}_{\PBH,i}(t)\right] \ .
\end{equation}
Here $m_\psi$ is the ULDM particle mass, $\MPBH$ is the individual PBH mass, and $\NPBH$ is the number of PBHs.
The PBH trajectories obey Newtonian equations in the gravitational potential generated by the ULDM field and by the other PBHs,
\begin{equation}
    \ddot{\bm{x}}_{\PBH,i}
    =
    -\nabla\Phi_{\mathrm{ext}}^{(i)}(t,\bm{x})\big|_{\bm{x}=\bm{x}_{\PBH,i}} \ ,
\end{equation}
where the external potential acting on the $i$th PBH is
\begin{equation}
    \Phi_{\mathrm{ext}}^{(i)}(t,\bm{x})
    =
    \Phi_\psi(t,\bm{x})
    +
    \sum_{j\ne i}^{\NPBH}\Phi_{\PBH,j}(t,\bm{x}) \ .
\end{equation}
These equations define the full dynamical problem.
In principle, they can be solved directly with a time-dependent Schr\"odinger-Poisson solver coupled to discrete PBH trajectories.
Rather than pursuing such direct simulations here, we decompose the system into a time-independent background Hamiltonian and fluctuations around it.
This separation isolates the equilibrium halo structure from the wave-interference and discrete PBH shot-noise fluctuations, which can then be treated statistically.

We decompose the total density and gravitational potential as
\begin{equation}
    \rho(t,\bm{x}) = \bar{\rho}(\bm{x})+\delta\rho(t,\bm{x}) \ ,
    \qquad
    \Phi(t,\bm{x}) = \bar{\Phi}(\bm{x})+\delta\Phi(t,\bm{x}) \ .
\end{equation}
Here barred quantities denote the background obtained after an appropriate time, phase, or ensemble average, and are related by the background Poisson equation,
\begin{equation}
    \label{eq:bg-poisson}
    \nabla^2\bar{\Phi} = 4\pi G\bar{\rho} \ .
\end{equation}
Throughout this work, we use the boundary condition $\Phi(t,\infty)=0$, and hence $\bar{\Phi}(\infty)=\delta\Phi(t,\infty)=0$.
At leading order, the ULDM eigenmodes are constructed from the static Hamiltonian through
\begin{equation}
    \bar{H}\psi_j(\bm{x})
    =
    E_j\psi_j(\bm{x}) \ ,
    \label{eq:bg-hamiltonian-eigenbasis}
\end{equation}
where
\begin{equation}
    \bar{H}
    =
    -\frac{\hbar^2}{2m_\psi}\nabla^2
    +m_\psi\bar{\Phi}(\bm{x}) \ ,
\end{equation}
while $\delta\Phi$ is treated as a perturbation.
The ULDM field can then be expanded as
\begin{equation}
    \psi(t,\bm{x})
    =
    \sum_j
    a_j(t)
    \psi_j(\bm{x})
    \ee^{-\ii E_jt/\hbar} \ .
    \label{eq:general-eigenmode-expansion}
\end{equation}

For the mixed ULDM--PBH system, the background density is decomposed into ULDM and PBH contributions,
\begin{equation}
    \bar{\rho}(\bm{x})
    =
    \bar{\rho}_\psi(\bm{x})
    +
    \bar{\rho}_{\PBH}(\bm{x}) \ ,
\end{equation}
and the integrated PBH abundance is
\begin{equation}
    \fPBH \equiv \frac{\int \dif^3x\,\bar{\rho}_{\PBH}(\bm{x})}{\int \dif^3x\,\bar{\rho}(\bm{x})} \ .
\end{equation}
The fluctuation around this background contains both ULDM wave-interference and discrete PBH shot-noise contributions,
\begin{equation}
    \delta\rho(t,\bm{x})
    =
    \delta\rho_\psi(t,\bm{x})
    +
    \delta\rho_{\PBH}(t,\bm{x}) \ ,
\end{equation}
where
\begin{equation}
    \rho_\psi(t,\bm{x})
    =
    \bar{\rho}_\psi(\bm{x})
    +
    \delta\rho_\psi(t,\bm{x}) \ ,
    \qquad
    \rho_{\PBH}(t,\bm{x})
    =
    \bar{\rho}_{\PBH}(\bm{x})
    +
    \delta\rho_{\PBH}(t,\bm{x}) \ ,
\end{equation}
which gives
\begin{equation}
    \left\langle \delta\rho_\psi \right\rangle = 0 \ ,
    \qquad
    \left\langle \delta\rho_{\PBH} \right\rangle = 0 \ .
    \label{eq:zero-mean-fluctuations}
\end{equation}
The averaging brackets should be interpreted according to the component being averaged.
For ULDM, the relevant average is a time or random-phase average that removes interference terms between distinct eigenmodes.
For PBHs, the brackets denote an average over discrete realizations, or an equivalent orbital-phase or coarse-grained average.
In this work we focus on subdominant PBH abundances, $\fPBH\ll1$, so that the background remains dominated by ULDM.

\section{Continuum PBH limit}
\label{sec:continuum-pbh-limit}

We first consider the continuum limit $\MPBH\to0$ and $\NPBH\to\infty$ at fixed background PBH density.
In this limit, the PBH population is described by a continuum density profile and $\delta\rho_{\PBH}\to0$.
The problem then reduces to constructing ULDM eigenmodes in a static background potential sourced by the mixed ULDM--PBH background density.
At the static background level, the continuum effect is not specific to PBHs: a collisionless particle-dark-matter component with the same density profile produces the same background potential and ULDM eigenproblem.

\subsection{Fiducial mixed halo}

We restrict the background to spherical symmetry, so all background profiles depend only on $r=|\bm{x}|$, and adopt the ULDM halo profile from the soliton--envelope construction used in Ref.~\cite{Yavetz:2021pbc}.
In this construction, the inner Navarro--Frenk--White (NFW) cusp is smoothly replaced by a soliton,
\begin{equation}
    \rho_{\mathrm{sNFW}}(r)
    =
    W_t(r)\rho_\sol(r)
    +
    [1-W_t(r)]\rho_\NFW(r) \ ,
    \qquad
    W_t(r)
    =
    \frac{1}{2}
    \left[
    1
    -
    \tanh\left(
    \frac{\ln(r/r_t)}{\Delta_t}
    \right)
    \right] \ ,
\end{equation}
where $r_t$ is the transition radius between the soliton core and the outer NFW-like envelope, defined by matching the soliton and NFW profiles $\rho_\sol(r_t)=\rho_\NFW(r_t)$, and $\Delta_t$ controls the transition width in logarithmic radius.
The continuum PBH component is not assumed to follow the soliton in the central region.
Ref.~\cite{Schwabe:2020eac} studied mixed halos of fuzzy dark matter and CDM, and found that the collisionless component does not share the solitonic central profile of the wave component, while the outer halo remains approximately NFW-like on scales larger than the de Broglie wavelength.
Motivated by this behavior, we model the continuum PBH component as a CDM-like tracer with an NFW outer profile and a finite-density central core,
\begin{equation}
    \rho_{\mathrm{cNFW}}(r)
    =
    A_{\mathrm{cNFW}}
    \frac{r}{r+r_t}
    \rho_\NFW(r) \ .
    \label{eq:cored-nfw-profile}
\end{equation}
Here the factor $r/(r+r_t)$ regularizes this profile on the transition scale $r_t$, and the normalization factor $A_{\mathrm{cNFW}}$ is fixed by requiring $\int \dif^3x\,\rho_{\mathrm{cNFW}}(r)=\int \dif^3x\,\rho_{\mathrm{sNFW}}(r)\equiv M_{\mathrm{tot}}$, once a common outer normalization radius $r_{\mathrm{norm}}$ is specified.
The resulting mass-normalized cored-NFW template keeps the PBH component finite in the central region, does not force it to trace the ULDM soliton, and recovers the NFW-like behavior at large radii.
The total background density is then
\begin{equation}
    \bar{\rho}(r)
    =
    \bar{\rho}_\psi(r)+\bar{\rho}_{\PBH}(r)
    =
    (1-\fPBH)\rho_{\mathrm{sNFW}}(r)
    +
    \fPBH\rho_{\mathrm{cNFW}}(r) \ .
    \label{eq:smooth-total-density}
\end{equation}
Together with the background Poisson equation, the background density fixes the spherical background potential $\bar{\Phi}(r)$, which defines the ULDM eigenmode basis.

We use a fiducial benchmark halo with the following parameters.
The outer NFW profile is~\cite{Navarro:1996gj}
\begin{equation}
    \rho_\NFW(r)
    =
    \frac{\rho_s}{(r/r_s)(1+r/r_s)^2}
    \exp\left(-\frac{r^2}{2r_{\mathrm{vir}}^2}\right) \ ,
\end{equation}
with $r_s=10\kpc$, $\rho_s=10^6\Msun/\kpc^3$, and $r_{\mathrm{vir}}=56\kpc$, corresponding to a halo mass of order $M_{\mathrm{vir}}\simeq10^{10}\Msun$.
Here the exponential factor is added to keep the total NFW mass finite.

\begin{figure}[htbp]
    \centering
    \includegraphics{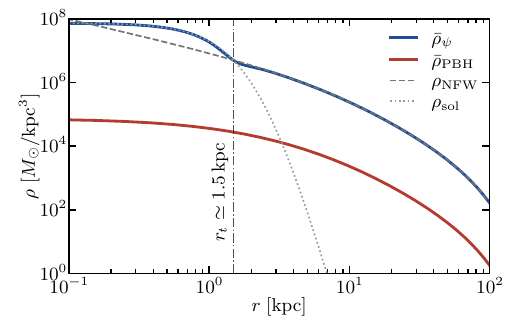}
    \caption{%
        Background density decomposition for the fiducial mixed halo.
        The blue and red curves show the background ULDM density, $\bar{\rho}_\psi=(1-\fPBH)\rho_{\mathrm{sNFW}}$, and the background PBH density, $\bar{\rho}_{\PBH}=\fPBH\rho_{\mathrm{cNFW}}$, whose sum defines the total background density $\bar{\rho}$.
        The gray dashed and dotted curves show the NFW envelope and soliton profile used to construct the hybrid templates.
        The vertical dash-dotted line marks the transition radius $r_t$.
    }
    \label{fig:density-profiles}
\end{figure}

The soliton profile is approximated by~\cite{Schive:2014dra}
\begin{equation}
    \rho_\sol(r)
    =
    \frac{
    1.9\times10^7
    (m_\psi/10^{-22}\eV)^{-2}
    (r_c/\kpc)^{-4}
    }{
    [1+0.091(r/r_c)^2]^8
    }
    \frac{\Msun}{\kpc^3} \ ,
\end{equation}
where the core radius is~\cite{Schive:2014hza}
\begin{equation}
    r_c
    =
    1.6
    \left(\frac{m_\psi}{10^{-22}\eV}\right)^{-1}
    \left(\frac{M_{\mathrm{vir}}}{10^9\Msun}\right)^{-1/3}
    \kpc \ .
\end{equation}
For the fiducial benchmark we take $m_\psi=10^{-22}\eV$, which gives $r_c\simeq0.7\kpc$ for $M_{\mathrm{vir}}\simeq10^{10}\Msun$.
With these parameters the matching condition gives $r_t\simeq1.5\kpc$, corresponding to $r_t/r_c\simeq2.1$.
We use a smooth transition width $\Delta_t=0.3$ and a normalization radius $r_{\mathrm{norm}}=6r_{\mathrm{vir}}=336\kpc$.
For the fiducial mixed benchmark we set $\fPBH=0.01$, so that one percent of the dark matter mass is assigned to the PBH component.
This value should be read as a fiducial value for the mixed-halo benchmark, since observational limits on PBH abundance depend strongly on the individual PBH mass~\cite{Carr:2020gox}.
Figure~\ref{fig:density-profiles} shows the corresponding density decomposition, together with the soliton and NFW profiles used to define the hybrid templates.

\subsection{Background eigenmodes}

\begin{figure}[htbp]
    \centering
    \includegraphics{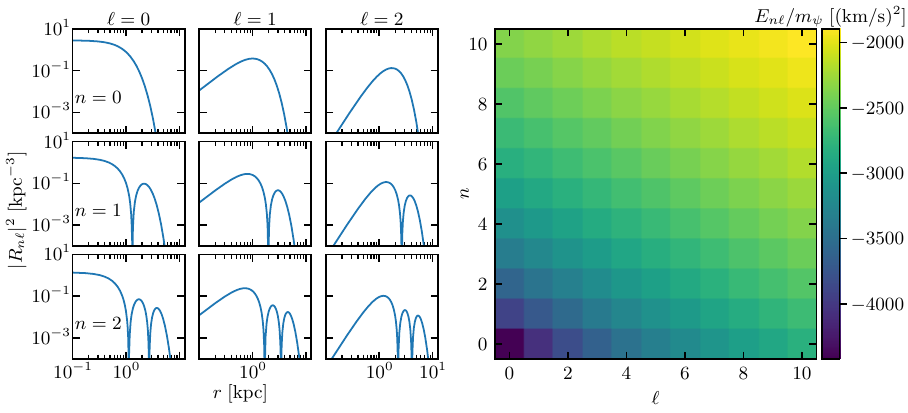}
    \caption{%
        Low-lying eigenmodes of the fiducial mixed halo.
        The radial profiles show the $\qm$-independent squared radial eigenfunctions $|R_{\qn\ql}(r)|^2$ for $0\leq\qn,\ql\leq2$.
        The accompanying spectrum shows the specific eigenvalues $E_{\qn\ql}/m_\psi$ for $0\leq\qn,\ql\leq10$.
    }
    \label{fig:eigenmode}
\end{figure}

For the fiducial mixed halo, we solve the resulting spherical eigenvalue problem.
The eigenmodes are labeled by $j=(\qn,\ql,\qm)$ and factorized as
\begin{equation}
    \psi_{\qn\ql\qm}(\bm{x})
    =
    R_{\qn\ql}(r)Y_{\ql\qm}(\hat{\bm{x}}) \ .
\end{equation}
With $u_{\qn\ql}(r)\equiv rR_{\qn\ql}(r)$, the background eigenvalue problem becomes
\begin{equation}
    \label{eq:fiducial-radial-eigenproblem}
    \left[
    -\frac{\hbar^2}{2m_\psi}\frac{\dif^2}{\dif r^2}
    +
    \frac{\hbar^2\ql(\ql+1)}{2m_\psi r^2}
    +
    m_\psi\bar{\Phi}(r)
    \right]
    u_{\qn\ql}(r)
    =
    E_{\qn\ql}u_{\qn\ql}(r) \ .
\end{equation}
By spherical symmetry, the eigenvalue is independent of quantum number $\qm$, so each $(\qn,\ql)$ radial solution represents $2\ql+1$ degenerate angular modes.
We normalize each eigenmode to unity,
\begin{equation}
    \label{eq:fiducial-radial-normalization}
    \int \dif^3x\,|\psi_{\qn\ql\qm}(\bm{x})|^2
    =
    \int_0^\infty \dif r\,r^2|R_{\qn\ql}(r)|^2
    =
    \int_0^\infty \dif r\,|u_{\qn\ql}(r)|^2
    =
    1 \ .
\end{equation}
Numerically, the radial eigenvalue problem is solved on a finite radial domain $0<r<r_{\mathrm{cut}}$ with $r_{\mathrm{cut}}=2.5r_{\mathrm{vir}}=140\kpc$ and Dirichlet boundary conditions for $u_{\qn\ql}$ at both endpoints.
Figure~\ref{fig:eigenmode} shows representative low-lying eigenmodes of the radial eigenvalue problem.
The left panel displays squared radial profiles for $0\leq\qn,\ql\leq2$, while the right panel shows the corresponding low-lying spectrum for $0\leq\qn,\ql\leq10$.

\begin{figure}[htbp]
    \centering
    \includegraphics{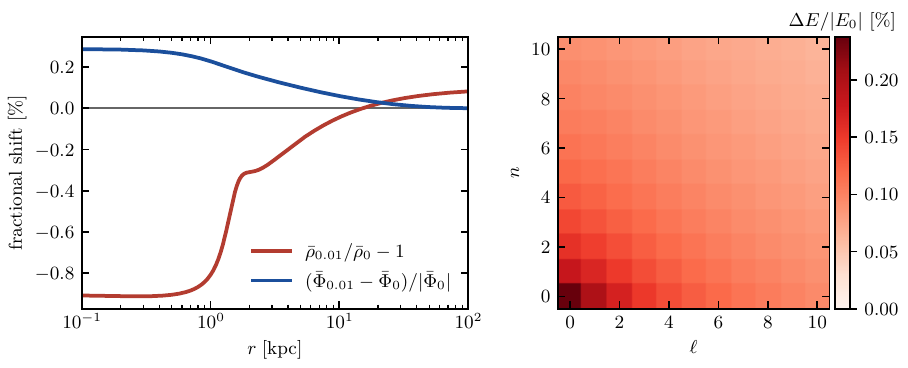}
    \caption{%
        Background-level shifts induced by the PBH abundance $\fPBH=0.01$, relative to the pure-ULDM case.
        Left: fractional change in the total background density and the corresponding change in the gravitational potential.
        Right: fractional eigenvalue shifts for the low-lying eigenmodes, with $\Delta E=E_{0.01}-E_0$.
    }
    \label{fig:smooth-pbh-mean}
\end{figure}

Varying $\fPBH$ changes $\bar{\rho}$ and $\bar{\Phi}$ at fixed total halo mass, and therefore changes the eigenvalues and eigenmodes.
We isolate the PBH effect at the background level by comparing the $\fPBH=0.01$ model to the pure-ULDM case $\fPBH=0$.
Figure~\ref{fig:smooth-pbh-mean} shows that the PBH component produces a sub-percent reshaping of the total density profile and a correspondingly small potential shift.
The associated low-lying eigenvalue shifts remain below about $0.25\%$ for the modes shown.

\subsection{Mode-amplitude fit}

The time-dependent ULDM field can be expanded as
\begin{equation}
    \psi(t,\bm{x})
    =
    \sum_j
    a_j
    \psi_j(\bm{x})
    \ee^{-\ii E_jt/\hbar} \ .
\end{equation}
Here the constant amplitudes $a_j$ reconstruct only the ULDM component and are chosen so that the averaged ULDM density matches the target profile,
\begin{equation}
    \left\langle \rho_\psi(t,\bm{x}) \right\rangle
    =
    m_\psi
    \sum_j
    |a_j|^2
    |\psi_j(\bm{x})|^2
    \simeq
    \bar{\rho}_{\psi}(r) \ .
\end{equation}
The PBH component is not part of the density reconstructed by the ULDM amplitudes, but it affects the amplitude fit through the eigenfunctions, since the eigenmodes depend on the background potential.

We take $|a_{\qn\ql\qm}|$ to be independent of $\qm$ within each $(\qn,\ql)$ multiplet, so the reconstructed radial density becomes
\begin{equation}
    \bar{\rho}_{\psi}^{\mathrm{fit}}(r)
    =
    m_\psi
    \sum_{\qn,\ql}
    \frac{2\ql+1}{4\pi}
    |a_{\qn\ql\qm}|^2
    |R_{\qn\ql}(r)|^2 \ .
\end{equation}
The corresponding fitted multiplet mass, $M_{\qn\ql}=m_\psi(2\ql+1)|a_{\qn\ql\qm}|^2$, is the ULDM mass carried by the $(\qn,\ql)$ multiplet.
Thus $\sum_{\qn,\ql}M_{\qn\ql}$ gives the total fitted ULDM mass, which approximates $(1-\fPBH)M_{\mathrm{tot}}$.
The amplitudes are obtained by minimizing the fractional radial mismatch~\cite{Yavetz:2021pbc},
\begin{equation}
    D
    =
    \frac{1}{r_{\mathrm{fit}}}
    \int_0^{r_{\mathrm{fit}}}
    \dif r\,
    \left[
    \frac{\bar{\rho}_{\psi}^{\mathrm{fit}}(r)-\bar{\rho}_{\psi}(r)}
    {\bar{\rho}_{\psi}(r)}
    \right]^2 \ .
\end{equation}

\begin{figure}[htbp]
    \centering
    \includegraphics{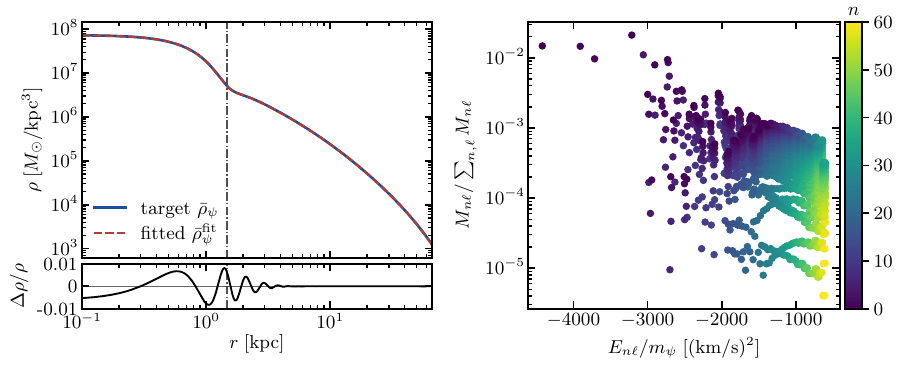}
    \caption{%
        Eigenmode-amplitude fit for the fiducial mixed halo.
        Left: target ULDM density $\bar{\rho}_\psi=(1-\fPBH)\rho_{\mathrm{sNFW}}$ and reconstructed density $\bar{\rho}_{\psi}^{\mathrm{fit}}$.
        The lower left subpanel shows the fractional residual $(\bar{\rho}_{\psi}^{\mathrm{fit}}-\bar{\rho}_\psi)/\bar{\rho}_\psi$ over the fitted range.
        The vertical dash-dotted line marks $r_t$.
        Right: fitted mass fraction $M_{\qn\ql}/\sum_{\qn,\ql}M_{\qn\ql}$ for populated $(\qn,\ql)$ multiplets with mass fraction above $10^{-12}$, plotted against the specific eigenvalue $E_{\qn\ql}/m_\psi$ and colored by $\qn$.
    }
    \label{fig:amplitude-fit}
\end{figure}

For the amplitude fit, we treat each $(\qn,\ql)$ multiplet as an independent nonnegative degree of freedom and include a regularization term favoring an isotropic energy-bin prior, with 100 energy bins and a regularization strength of $10^{-5}$.
This prescription suppresses sparse solver artifacts in random-phase realizations while retaining the freedom to assign different masses to different angular momenta at fixed energy.
In the fit, we use $r_{\mathrm{fit}}=1.2r_{\mathrm{vir}}$ and select modes whose energies lie below the circular-orbit energy at $r_{\mathrm{vir}}$.
The amplitude-fit results give $\sqrt{D}\simeq9.1\times10^{-4}$ over the fitted range and assign positive mass to 1965 of the 2013 selected multiplets, with 1938 having fitted mass fractions $M_{\qn\ql}/\sum_{\qn,\ql}M_{\qn\ql}>10^{-12}$.
Figure~\ref{fig:amplitude-fit} shows the resulting density reconstruction and the fitted mass carried by each populated eigenmode multiplet.
The small residual in the density panel demonstrates that this finite eigenbasis is sufficient for the target profile over the fitted range, and the mode-mass panel shows that a broad set of $(\qn,\ql)$ multiplets contributes to the reconstruction.

\subsection{ULDM granularity}

\begin{figure}[htbp]
    \centering
    \includegraphics{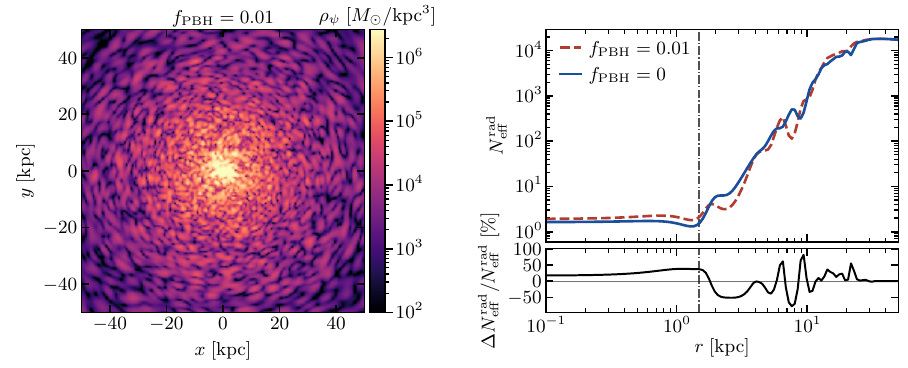}
    \caption{%
        Granular ULDM density slice and radial participation measure.
        Left: representative ULDM density slice at a fixed time over $-50\kpc\le x,y\le50\kpc$ for the fiducial mixed halo.
        Upper right: radial participation measure $N_{\mathrm{eff}}^{\mathrm{rad}}(r)$ for the mixed and pure-ULDM fits.
        Lower right: fractional shift $N_{\mathrm{eff}}^{\mathrm{rad}}(\fPBH=0.01)/N_{\mathrm{eff}}^{\mathrm{rad}}(\fPBH=0)-1$.
        The vertical dash-dotted line marks the soliton--envelope transition radius $r_t$.
    }
    \label{fig:random-granularity}
\end{figure}

With the mode amplitudes fixed, the phase ensemble is specified by assigning an independent random phase drawn uniformly from $[0,2\pi)$ to each eigenmode,
\begin{equation}
    a_{\qn\ql\qm}
    =
    |a_{\qn\ql\qm}|
    \ee^{\ii\theta_{\qn\ql\qm}} \ ,
    \qquad
    \theta_{\qn\ql\qm}\sim U(0,2\pi) \ .
\end{equation}
The density is then
\begin{equation}
    \rho_\psi(t,\bm{x})
    =
    m_\psi|\psi(t,\bm{x})|^2
    =
    m_\psi\sum_j |a_j|^2|\psi_j(\bm{x})|^2
    +
    m_\psi\sum_{j\ne k}
    a_j a_k^*
    \psi_j(\bm{x})\psi_k^*(\bm{x})
    \ee^{-\ii(E_j-E_k)t/\hbar} \ .
\end{equation}
It contains the fitted diagonal profile and interference terms between distinct eigenmodes.
These interference terms generate the granular ULDM density pattern shown in the left panel of Fig.~\ref{fig:random-granularity}.

At the background level, the continuum PBH component affects the background potential, eigenbasis, and fitted mode weights, and therefore changes the statistics of the granular field.
The phase-ensemble mean is fixed by construction to the fitted background profile,
\begin{equation}
    \left\langle \rho_\psi(\bm{x}) \right\rangle_\theta
    =
    \bar{\rho}_{\psi}^{\mathrm{fit}}(r)
    \simeq
    \bar{\rho}_{\psi}(r) \ ,
\end{equation}
where $\langle\cdots\rangle_\theta$ denotes an ensemble average over the independent random phases.
We characterize the phase ensemble at a fixed time with the normalized density fluctuation
\begin{equation}
    \delta_\psi(\bm{x})
    =
    \frac{
    \rho_\psi(\bm{x})
    -
    \left\langle \rho_\psi(\bm{x}) \right\rangle_\theta
    }{
    \left\langle \rho_\psi(\bm{x}) \right\rangle_\theta
    } \ ,
\end{equation}
which has the local variance
\begin{equation}
    \operatorname{var}(\delta_\psi(\bm{x}))
    =
    \left\langle \delta_\psi^2(\bm{x}) \right\rangle_\theta
    =
    1
    -
    \frac{1}{N_{\mathrm{eff}}(\bm{x})} \ ,
    \label{eq:smooth-limit-granularity-variance}
\end{equation}
where the local participation ratio is
\begin{equation}
    N_{\mathrm{eff}}(\bm{x})
    =
    \frac{
    \left[
    \sum_j
    |a_j|^2
    |\psi_j(\bm{x})|^2
    \right]^2
    }{
    \sum_j
    |a_j|^4
    |\psi_j(\bm{x})|^4
    } \ .
\end{equation}
The local participation ratio $N_{\mathrm{eff}}$ measures the effective number of modes contributing locally.
Small $N_{\mathrm{eff}}$ corresponds to coherent few-mode interference, while large $N_{\mathrm{eff}}$ approaches the many-mode random-wave limit with $\sqrt{\operatorname{var}(\delta_\psi)}\simeq1$.
A one-dimensional radial participation measure is obtained by angular averaging of the individual $\qm$ contributions,
\begin{equation}
    N_{\mathrm{eff}}^{\mathrm{rad}}(r)
    =
    \frac{
    \left[
    \sum_{\qn,\ql}
    (2\ql+1)
    \frac{|a_{\qn\ql\qm}|^2 |R_{\qn\ql}(r)|^2}{4\pi}
    \right]^2
    }{
    \sum_{\qn,\ql}
    (2\ql+1)
    \left[
    \frac{|a_{\qn\ql\qm}|^2 |R_{\qn\ql}(r)|^2}{4\pi}
    \right]^2
    } \ .
\end{equation}
The right panels of Fig.~\ref{fig:random-granularity} compare $N_{\mathrm{eff}}^{\mathrm{rad}}(r)$ and its fractional shift for $\fPBH=0.01$ and $\fPBH=0$.
Within the soliton--envelope transition radius, $r\lesssim r_t\simeq1.5\kpc$, the continuum PBH component coherently increases the radial participation, with a median enhancement of about $20\%$.
For $r\gtrsim r_t$, the fractional response becomes oscillatory around a median close to zero and decays toward the outer halo, while the phase-ensemble rms density contrast remains nearly unchanged.
Together, these trends show that the coherent inner ULDM mode mixture is sensitive to a percent-level continuum PBH component at the tens-of-percent level, while the outer halo remains close to the many-mode random-wave regime.
This contrast arises because the inner soliton region is formed from a relatively coherent few-mode ULDM mixture, whereas the outer NFW-like envelope contains many radially excited ULDM modes whose small PBH-induced shifts mostly average out.

\section{Discrete PBH effects}
\label{sec:discrete-pbh-effects}

The previous section studied the mixed ULDM--PBH halo in the continuum limit $\MPBH\to0$, thereby isolating the background-level effect of the PBH component.
For nonzero individual PBH mass, the continuum PBH background density is represented by a finite number of compact objects and therefore produces shot-noise fluctuations due to the discrete nature of the PBH distribution.
PBHs can acquire dark-matter minihalos before entering a galactic halo in suitable parameter regimes~\cite{Hertzberg:2020hsz,Yin:2024xov}.
Efficient minihalo accretion requires the dark-matter de Broglie wavelength $\lambda_{\mathrm{dB}}$ to be smaller than the minihalo size $R_{\mathrm{halo}}$; when $\lambda_{\mathrm{dB}}/R_{\mathrm{halo}}\gtrsim1$, the wave nature of dark matter cannot be neglected and accretion becomes inefficient~\cite{Hertzberg:2020hsz}.
For $m_\psi=10^{-22}\eV$, $\MPBH=1\Msun$, and redshift $z=30$, the corresponding estimate gives $\lambda_{\mathrm{dB}}/R_{\mathrm{halo}}\simeq1.0\times10^5$.
Thus, the accretion required to build such a minihalo is strongly wave-suppressed in our fiducial stellar-mass benchmark.
We therefore model the PBHs as bare compact perturbers and neglect any ULDM dressing bound to individual PBHs in the calculation below.
Accordingly, we decompose the discrete PBH density as
\begin{equation}
    \rho_{\PBH}(t,\bm{x})
    =
    \sum_{i=1}^{\NPBH}
    \MPBH\delta^{(3)}\!\left[\bm{x}-\bm{x}_{\PBH,i}(t)\right]
    =
    \bar{\rho}_{\PBH}(\bm{x})
    +
    \delta\rho_{\PBH}(t,\bm{x}) \ .
\end{equation}
Here $\delta\rho_{\PBH}$ denotes the residual relative to the background PBH profile obtained by the corresponding ensemble, orbital-phase, or coarse-grained average, and we have $\langle\delta\rho_{\PBH}\rangle=0$.

\subsection{PBH shot noise}

\begin{figure}[htbp]
    \centering
    \includegraphics{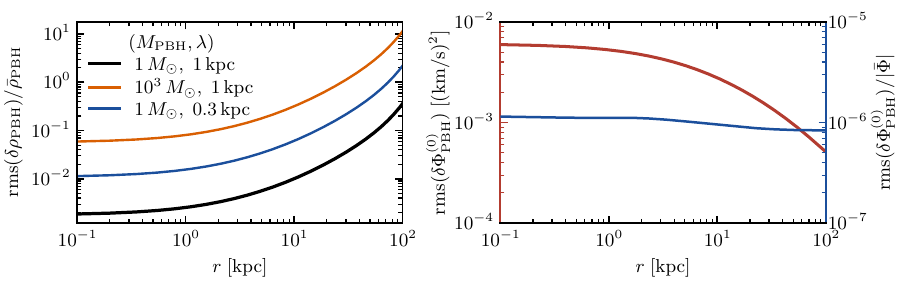}
    \caption{%
        PBH shot-noise estimates for the density and potential fluctuations in the fiducial mixed halo.
        Left: local fractional density fluctuation from Eq.~\eqref{eq:fractional-rms-density-fluctuation}.
        Right: monopole potential fluctuation $\operatorname{rms}(\delta\Phi_{\PBH}^{(0)})=\sqrt{\operatorname{var}(\delta\Phi_{\PBH}^{(0)})}$ from Eq.~\eqref{eq:pbh-monopole-potential-variance}, and its ratio to the background potential.
    }
    \label{fig:pbh-density-potential-fluctuation}
\end{figure}

For a region of volume $V$ over which $\bar{\rho}_{\PBH}$ is approximately constant, the PBH count can be approximated by a Poisson distribution~\cite{Afshordi:2003zb}
\begin{equation}
    N_{\PBH}(V)
    \sim
    \mathrm{Poisson}(\bar{N}_{\PBH}) \ ,
    \qquad
    \bar{N}_{\PBH}
    =
    \frac{\bar{\rho}_{\PBH}V}{\MPBH} \ ,
\end{equation}
with variance
\begin{equation}
    \operatorname{var}(N_{\PBH})
    =
    \bar{N}_{\PBH} \ .
\end{equation}
The same Poisson statistics give the coarse-grained density variance in this region,
\begin{equation}
    \operatorname{var}(\rho_{\PBH})
    =
    \frac{\MPBH^2}{V^2}\operatorname{var}(N_{\PBH})
    =
    \frac{\MPBH\bar{\rho}_{\PBH}}{V} \ ,
\end{equation}
which gives the corresponding fractional rms density fluctuation,
\begin{equation}
    \frac{\operatorname{rms}(\delta\rho_{\PBH})}{\bar{\rho}_{\PBH}}
    =
    \frac{\sqrt{\operatorname{var}(\delta\rho_{\PBH})}}{\bar{\rho}_{\PBH}}
    =
    \frac{\sqrt{\operatorname{var}(\rho_{\PBH})}}{\bar{\rho}_{\PBH}}
    =
    \left(\frac{\MPBH}{\bar{\rho}_{\PBH}V}\right)^{1/2}
    =
    \frac{1}{\sqrt{\bar{N}_{\PBH}}} \ .
    \label{eq:fractional-rms-density-fluctuation}
\end{equation}
At fixed PBH background profile $\bar{\rho}_{\PBH}$, this shot-noise fluctuation scales as $\MPBH^{1/2}$.
Therefore, the fluctuation is larger for larger individual PBH masses, and it vanishes in the continuum limit.
For the local estimate, we take a spherical top-hat smoothing volume $V=4\pi\lambda^3/3$ with radius $\lambda$, and the left panel of Fig.~\ref{fig:pbh-density-potential-fluctuation} shows the resulting $\MPBH^{1/2}$ and $\lambda^{-3/2}$ scalings.

The associated gravitational potential fluctuation is determined by Poisson's equation,
\begin{equation}
    \nabla^2\delta\Phi_{\PBH}
    =
    4\pi G\delta\rho_{\PBH} \ .
\end{equation}
Under the boundary condition $\Phi(t,\infty)=0$, the potential fluctuation is well defined relative to spatial infinity.
To estimate the PBH potential fluctuation, we consider its monopole component.
A thin shell at radius $s$ with fluctuating shell mass $\dif\delta M_{\PBH}(t;s)$ contributes
\begin{equation}
    \dif\delta\Phi_{\PBH}^{(0)}(t,r;s)
    =
    -\frac{G\,\dif\delta M_{\PBH}(t;s)}{\max(r,s)} \ ,
    \label{eq:pbh-shell-monopole-potential}
\end{equation}
where the superscript $^{(0)}$ denotes the monopole component.
The denominator follows from the shell theorem: outside the shell, the fluctuation is the point-mass potential $-G\dif\delta M_{\PBH}(t;s)/r$, while inside the shell it is the constant potential $-G\dif\delta M_{\PBH}(t;s)/s$.
Adding the contributions from independent radial shells, we obtain the variance of the monopole PBH potential fluctuation,
\begin{equation}
    \begin{aligned}
    \operatorname{var}\!\left(\delta\Phi_{\PBH}^{(0)}(t,r)\right)
    &=
    \int_0^{r_{\mathrm{cut}}}
    \operatorname{var}\!\left(\dif\delta\Phi_{\PBH}^{(0)}(t,r;s)\right)
    =
    \int_0^{r_{\mathrm{cut}}}
    \frac{G^2 \operatorname{var}\!\left(\dif\delta M_{\PBH}(t;s)\right)}{\max(r,s)^2}
    \\
    &=
    G^2\MPBH
    \int_0^{r_{\mathrm{cut}}}
    \frac{4\pi s^2\dif s\,\bar{\rho}_{\PBH}(s)}{\max(r,s)^2} \ ,
    \end{aligned}
    \label{eq:pbh-monopole-potential-variance}
\end{equation}
where the last equality uses the shell mass variance for a shell at radius $s$ with independent PBH number fluctuations,
\begin{equation}
    \operatorname{var}\!\left(\dif\delta M_{\PBH}(t;s)\right)
    =
    \MPBH^2\operatorname{var}\!\left(\dif\delta N_{\PBH}(t;s)\right)
    =
    \MPBH^2\dif\bar{N}_{\PBH}(s)
    =
    \MPBH\,4\pi s^2\dif s\,\bar{\rho}_{\PBH}(s) \ .
    \label{eq:pbh-shell-mass-variance}
\end{equation}
The monopole potential variance gives the spherically averaged contribution of PBH shot noise, yielding a one-dimensional estimate of the gravitational potential fluctuation.
The right panel of Fig.~\ref{fig:pbh-density-potential-fluctuation} shows the estimate of monopole potential fluctuation and its ratio to the background potential, which is of order $10^{-6}$ for $\MPBH=1\Msun$.
These quantities characterize fluctuations in the depth of the potential well relative to spatial infinity, rather than fluctuations in the local gravitational acceleration.

\subsection{Monopole mode coupling}

The preceding subsection estimated the amplitude of the shot-noise fluctuation from the discrete PBH component.
This fluctuation acts as a time-dependent perturbation to the ULDM wave dynamics about the background,
\begin{equation}
    \ii\hbar\frac{\partial\psi}{\partial t}
    =
    \left[
    \bar{H}
    +
    m_\psi\delta\Phi_{\PBH}(t,\bm{x})
    \right]\psi
    +
    \cdots \ .
\end{equation}
Here the perturbation $\delta\Phi_{\PBH}$ denotes the PBH potential fluctuation, while the ellipsis represents other fluctuations, such as the intrinsic ULDM interference contribution to the gravitational potential.
To assess the dynamical effect of PBH fluctuations, we project the perturbation $\delta\Phi_{\PBH}$ onto the background eigenmodes, which gives the stochastic matrix elements driving transitions between modes.

We expand the ULDM field in eigenmodes of the background as
\begin{equation}
    \psi(t,\bm{x})
    =
    \sum_j
    a_j(t)\psi_j(\bm{x})
    \ee^{-\ii E_jt/\hbar} \ ,
\end{equation}
where $\psi_j$ and $E_j$ depend on the continuum PBH contribution to the background.
The PBH fluctuation contributes the perturbing Hamiltonian matrix elements
\begin{equation}
    V_{jk}(t)
    =
    m_\psi
    \int \dif^3x\,
    \psi_j^*(\bm{x})
    \delta\Phi_{\PBH}(t,\bm{x})
    \psi_k(\bm{x}) \ .
\end{equation}
By orthonormality, an additive spatially uniform shift of the potential contributes only to diagonal matrix elements and therefore cannot induce transitions between distinct modes.
At first order in the perturbation, the mode amplitudes obey
\begin{equation}
    \ii\hbar \dot{a}_j(t)
    =
    \sum_k
    V_{jk}(t)
    \ee^{\ii(E_j-E_k)t/\hbar}
    a_k(t) \ .
\end{equation}

The transition rate between two background eigenmodes is controlled by the spectral density of this stochastic matrix element.
For stationary noise we define
\begin{equation}
    S_{jk}(\omega)
    =
    \int \dif\tau\,
    \ee^{\ii\omega\tau}
    \left\langle
    V_{jk}(t)
    \left[
    V_{jk}(t-\tau)
    \right]^*
    \right\rangle \ .
\end{equation}
The leading-order transition estimate then gives
\begin{equation}
    \Gamma_{k\to j}
    \simeq
    \frac{1}{\hbar^2}
    S_{jk}(\omega_{jk}) \ ,
    \qquad
    \omega_{jk}
    =
    \frac{E_j-E_k}{\hbar} \ .
\end{equation}
This expression is the stochastic analogue of Fermi's golden rule: the PBH shot-noise fluctuation drives transitions between ULDM eigenmodes through the spectral power of the potential fluctuation at the transition frequency $\omega_{jk}$.

We first apply this transition-rate estimate to the monopole component of the PBH shot-noise fluctuation.
For $\ql=0$ modes with $u_{\qn0}(r)=rR_{\qn0}(r)$, the contribution of a fluctuating shell at radius $s$ to the monopole Hamiltonian matrix element is
\begin{equation}
    \begin{aligned}
    \dif V_{\qn\qn'}^{(0)}(t;s)
    &=
    m_\psi
    \int_0^{r_{\mathrm{cut}}}
    \dif r\,
    u_{\qn0}(r)
    \dif\delta\Phi_{\PBH}^{(0)}(t,r;s)
    u_{\qn'0}(r)
    \\
    &=
    -Gm_\psi\,\dif\delta M_{\PBH}(t;s)
    \int_0^{r_{\mathrm{cut}}}
    \dif r\,
    \frac{u_{\qn0}(r)u_{\qn'0}(r)}
    {\max(r,s)} \ ,
    \end{aligned}
\end{equation}
where Eq.~\eqref{eq:pbh-shell-monopole-potential} is used for the second equality.

We assign temporal correlations using a local PBH crossing time.
The PBH velocity dispersion is obtained from the isotropic Jeans equation for the PBH tracer in the mixed potential~\cite{2008gady.book.....B},
\begin{equation}
    \sigma_v^2(r)
    =
    \frac{1}{\bar{\rho}_{\PBH}(r)}
    \int_r^{r_{\mathrm{cut}}}
    \dif s\,
    \bar{\rho}_{\PBH}(s)
    \frac{\dif\bar{\Phi}}{\dif s} \ .
    \label{eq:pbh-jeans-sigma}
\end{equation}
We take the local correlation time to be
\begin{equation}
    \tau_c(r)
    =
    \frac{\lambda}{\sigma_v(r)} \ ,
\end{equation}
where $\lambda$ is the same top-hat coarse-graining scale used in the equal-time density-noise estimate above.
Under this local correlation model, each source shell is assigned an exponential autocorrelation with correlation time $\tau_c(s)$, and cross-shell correlations between distinct shells are neglected.
Combining the shell response with the shell mass variance in Eq.~\eqref{eq:pbh-shell-mass-variance} gives the unequal-time correlator
\begin{equation}
    \left\langle
    \dif V_{\qn\qn'}^{(0)}(t;s)
    \left[
    \dif V_{\qn\qn'}^{(0)}(t-\tau;s)
    \right]^*
    \right\rangle
    =
    G^2m_\psi^2\MPBH\,
    \mathcal{K}_{\qn\qn'}^{(0)}(s)
    \dif s\,
    \ee^{-|\tau|/\tau_c(s)} \ ,
\end{equation}
where the reduced radial kernel is
\begin{equation}
    \mathcal{K}_{\qn\qn'}^{(0)}(s)
    =
    4\pi s^2
    \bar{\rho}_{\PBH}(s)
    \left[
    \int_0^{r_{\mathrm{cut}}}
    \dif r\,
    \frac{u_{\qn0}(r)u_{\qn'0}(r)}
    {\max(r,s)}
    \right]^2 \ .
    \label{eq:pbh-monopole-mode-variance-kernel}
\end{equation}
The corresponding shell-integrated Hamiltonian matrix-element spectrum is then
\begin{equation}
    \begin{aligned}
    S_{\qn\qn'}^{(0)}(\omega)
    &=
    \int_{-\infty}^{\infty} \dif\tau\,
    \ee^{\ii\omega\tau}
    \int_0^{r_{\mathrm{cut}}}
    \left\langle
    \dif V_{\qn\qn'}^{(0)}(t;s)
    \left[
    \dif V_{\qn\qn'}^{(0)}(t-\tau;s)
    \right]^*
    \right\rangle
    \\
    &=
    G^2m_\psi^2\MPBH
    \int_0^{r_{\mathrm{cut}}}
    \dif s\,
    \frac{
    2\tau_c(s)
    }{
    1+\omega^2\tau_c^2(s)
    }
    \mathcal{K}_{\qn\qn'}^{(0)}(s) \ ,
    \end{aligned}
\end{equation}
where the Fourier transform of the exponential factor is used,
\begin{equation}
    \int_{-\infty}^{\infty}
    \dif\tau\,
    \ee^{\ii\omega\tau}
    \ee^{-|\tau|/\tau_c(s)}
    =
    \frac{2\tau_c(s)}{1+\omega^2\tau_c^2(s)} \ .
\end{equation}
This local exponential model of the temporal correlation fixes the equal-time PBH shot-noise amplitude and assigns a decorrelation time set by the Jeans velocity dispersion.
A more complete treatment would construct the time-dependent PBH density correlator from the PBH phase-space distribution and project it onto the mode-coupling kernels.
Such an orbit-based correlator would refine the spectral shape at the mode-splitting frequencies, but is not expected to change the rate estimates by many orders of magnitude.
The resulting transition-rate estimate for each mode pair $(\qn,\qn')$ is
\begin{equation}
    \Gamma_{\qn'\to\qn}^{(0)}
    \simeq
    \frac{1}{\hbar^2} S_{\qn\qn'}^{(0)}(\omega_{\qn\qn'}) \ ,
    \qquad
    \omega_{\qn\qn'}
    =
    \frac{E_{\qn0}-E_{\qn'0}}{\hbar} \ .
    \label{eq:pbh-transition-rate-estimate}
\end{equation}
At fixed $\fPBH$, the transition rates are proportional to $\MPBH$ because they are set by the variance of the PBH shot noise, and therefore the corresponding transition times scale as $\MPBH^{-1}$.

\begin{figure}[htbp]
    \centering
    \includegraphics{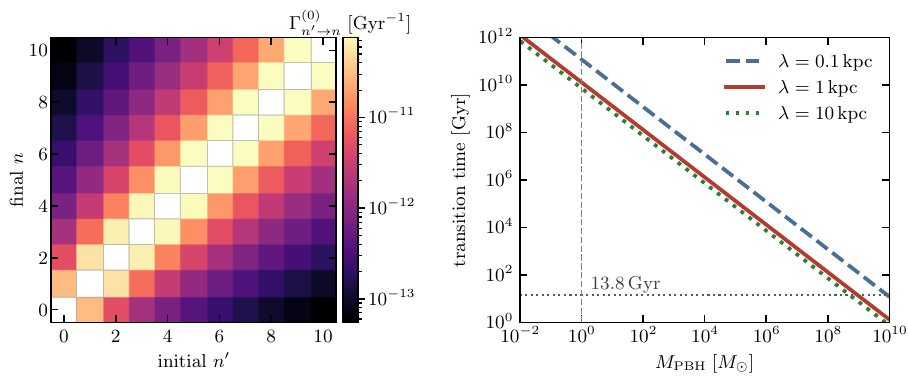}
    \caption{%
        Monopole PBH transition-rate estimate for the fiducial mixed halo.
        Left: transition rates between the first eleven $\ql=0$ radial modes, $\qn,\qn'=0,\ldots,10$, using Eq.~\eqref{eq:pbh-transition-rate-estimate} with $\MPBH=1\Msun$ and $\lambda=1\,\kpc$.
        The fastest transition is $\qn'=8\to\qn=7$ and has timescale $\simeq1.3\times10^{10}\,\mathrm{Gyr}$.
        Right: inverse-$\MPBH$ scaling of the transition time for $\lambda=0.1,1,10\,\kpc$.
        For each $\lambda$, the curve shows the fastest transition among the $11\times10=110$ directed off-diagonal transitions.
    }
    \label{fig:pbh-transition-rates}
\end{figure}

Figure~\ref{fig:pbh-transition-rates} shows the resulting monopole transition-rate estimate for the fiducial mixed halo.
The left panel shows the transition rates with $\MPBH=1\Msun$ and $\lambda=1\,\kpc$, and the fastest transition among the first eleven $\ql=0$ radial modes has a timescale of order $10^{10}\,\mathrm{Gyr}$.
Thus, for stellar-mass PBHs, monopole shot-noise-driven mode mixing is dynamically negligible for the low-lying ULDM modes.
This monopole calculation provides a transparent baseline for mode-pair transition rates driven by the PBH shot-noise fluctuation.
The right panel shows the inverse-$\MPBH$ scaling of the fastest transition time and the residual sensitivity to $\lambda$.
This scaling shows that the fastest monopole-channel transition would require $\MPBH\sim10^9\Msun$ to become comparable to the age of the Universe for $\lambda=1\,\kpc$.
These large-$\MPBH$ extrapolations should only be viewed as illustrative extensions of the perturbative estimate.
Varying the correlation length over two orders of magnitude, from $\lambda=0.1\,\kpc$ to $10\,\kpc$, changes the fastest monopole transition time by only a factor of about 16, which provides a sensitivity test of the local temporal-correlation model.
This weak sensitivity, combined with the fact that the transition times for stellar-mass PBHs lie far above the age of the Universe, implies that the negligible impact of stellar-mass PBH shot noise is robust to reasonable changes in the temporal-correlation model.

Mode transitions driven by PBH shot-noise fluctuations provide a channel for heating the central soliton region of the ULDM halo.
To estimate this effect, we define a soliton-region transition rate,
\begin{equation}
    \Gamma_{\mathrm{sol},\,\qn'\to\qn}^{(0)}
    \simeq
    \frac{1}{\hbar^2}
    S_{\mathrm{sol},\qn\qn'}^{(0)}(\omega_{\qn\qn'}) \ .
    \label{eq:pbh-soliton-region-rate}
\end{equation}
Here the soliton-region spectrum is obtained by restricting the source shells to the soliton--envelope transition radius $r_t$,
\begin{equation}
    S_{\mathrm{sol},\qn\qn'}^{(0)}(\omega)
    =
    G^2m_\psi^2\MPBH
    \int_0^{r_t}
    \dif s\,
    \frac{
    2\tau_c(s)
    }{
    1+\omega^2\tau_c^2(s)
    }
    \mathcal{K}_{\qn\qn'}^{(0)}(s) \ ,
    \label{eq:pbh-soliton-region-spectrum}
\end{equation}
which includes only the contribution from PBH fluctuations inside the soliton-dominated region.
For an initially ground-state-dominated soliton with occupation $N_0$, the associated heating-rate estimate is
\begin{equation}
    \dot{E}_{\mathrm{sol}}^{(0)}
    \simeq
    N_0
    \sum_{\qn>0}
    \left(E_{\qn0}-E_{00}\right)
    \Gamma_{\mathrm{sol},\,0\to\qn}^{(0)} \ .
    \label{eq:pbh-soliton-heating-rate}
\end{equation}
The corresponding heating time is then given by
\begin{equation}
    t_{\mathrm{heat}}^{(0)}
    \equiv
    \left|
    \frac{N_0E_{00}}{\dot{E}_{\mathrm{sol}}^{(0)}}
    \right|
    =
    \left|
    \frac{E_{00}}{
    \sum_{\qn>0}
    \left(E_{\qn0}-E_{00}\right)
    \Gamma_{\mathrm{sol},\,0\to\qn}^{(0)}
    }
    \right| \ ,
    \label{eq:pbh-soliton-heating-time}
\end{equation}
which is independent of $N_0$.
For the fiducial model with $\lambda=1\,\kpc$ and $\MPBH=1\Msun$, summing over $\ql=0$ modes up to $\qn=60$ gives a soliton-heating time of order $10^{11}\,\mathrm{Gyr}$.
Thus the soliton heating is negligible for stellar-mass PBHs in the fiducial mixed halo.
Because these monopole estimates omit the angular multipoles of the shot-noise potential, they provide a conservative baseline and motivate the multipole-resolved calculation in the next subsection.

\subsection{Angular multipole coupling}

We now include the angular multipoles of the PBH potential fluctuation and write the fluctuating PBH potential as
\begin{equation}
    \delta\Phi_{\PBH}(t,\bm{x})
    =
    \sum_{\mathfrak{l}\mathfrak{m}}
    \delta\Phi_{\mathfrak{l}\mathfrak{m}}(t,r)
    Y_{\mathfrak{l}\mathfrak{m}}(\hat{\bm{x}}) \ ,
\end{equation}
where $\mathfrak{l}$ and $\mathfrak{m}$ label the angular multipole and its azimuthal component.
The fluctuating shell mass multipole of a thin PBH-density shell at radius $s$ is
\begin{equation}
    \dif\delta M_{\mathfrak{l}\mathfrak{m}}(t;s)
    =
    s^2\dif s\,
    \int \dif\Omega\,
    Y_{\mathfrak{l}\mathfrak{m}}^*(\hat{\bm{x}})
    \delta\rho_{\PBH}(t,s,\hat{\bm{x}}) \ .
\end{equation}
With unit-normalized spherical harmonics and independent PBH Poisson fluctuations, the same-shell covariance is
\begin{equation}
    \left\langle
    \dif\delta M_{\mathfrak{l}\mathfrak{m}}(t;s)
    \dif\delta M_{\mathfrak{l}'\mathfrak{m}'}^*(t;s)
    \right\rangle
    =
    \MPBH\,
    s^2\dif s\,
    \bar{\rho}_{\PBH}(s)
    \delta_{\mathfrak{l}\mathfrak{l}'}
    \delta_{\mathfrak{m}\mathfrak{m}'} \ .
    \label{eq:pbh-multipole-shell-mass-covariance}
\end{equation}
Solving Poisson's equation for $\delta\Phi_{\PBH}$ with the isolated Newtonian Green's function gives the corresponding shell contribution to the potential multipole,
\begin{equation}
    \dif\delta\Phi_{\mathfrak{l}\mathfrak{m}}(t,r;s)
    =
    -\frac{4\pi G}{2\mathfrak{l}+1}
    \dif\delta M_{\mathfrak{l}\mathfrak{m}}(t;s)
    \frac{[\min(r,s)]^{\mathfrak{l}}}{[\max(r,s)]^{\mathfrak{l}+1}} \ .
\end{equation}

The contribution of this shell to the Hamiltonian matrix element between individual ULDM magnetic substates, labeled by $(\qn,\ql,\qm)$ and $(\qn',\ql',\qm')$, is
\begin{equation}
    \begin{aligned}
    &\dif V_{\qn\ql\qm;\qn'\ql'\qm'}^{\mathfrak{l}\mathfrak{m}}(t;s)
    \\
    &=
    m_\psi
    \int_0^{r_{\mathrm{cut}}}
    \dif r\,
    \int \dif\Omega\,
    \left[
    u_{\qn\ql}(r)Y_{\ql\qm}^*
    \right]
    \left[
    \dif\delta\Phi_{\mathfrak{l}\mathfrak{m}}(t,r;s)
    Y_{\mathfrak{l}\mathfrak{m}}
    \right]
    \left[
    u_{\qn'\ql'}(r)Y_{\ql'\qm'}
    \right]
    \\
    &=
    -Gm_\psi\,
    \dif\delta M_{\mathfrak{l}\mathfrak{m}}(t;s)
    \left[
    \frac{4\pi}{2\mathfrak{l}+1}
    \int_0^{r_{\mathrm{cut}}}
    \dif r\,
    u_{\qn\ql}(r)u_{\qn'\ql'}(r)
    \frac{[\min(r,s)]^{\mathfrak{l}}}{[\max(r,s)]^{\mathfrak{l}+1}}
    \right]
    \!
    \left[
    \int \dif\Omega\,
    Y_{\ql\qm}^*
    Y_{\mathfrak{l}\mathfrak{m}}
    Y_{\ql'\qm'}
    \right]
    \ .
    \end{aligned}
\end{equation}
Applying the same local temporal-correlation model as in the monopole estimate,
summing over the final $\qm$ states and source azimuthal multipoles $\mathfrak{m}$,
and averaging over the initially degenerate $\qm'$ states, which gives the factor $1/(2\ql'+1)$,
we obtain the multiplet-averaged unequal-time correlator,
\begin{equation}
    \begin{aligned}
    &\left\langle
    \dif V_{\qn\ql;\qn'\ql'}^{\mathfrak{l}}(t;s)
    \left[
    \dif V_{\qn\ql;\qn'\ql'}^{\mathfrak{l}}(t-\tau;s)
    \right]^*
    \right\rangle
    \\
    &\equiv
    \frac{1}{2\ql'+1}
    \sum_{\qm\,\mathfrak{m}\,\qm'}
    \left\langle
    \dif V_{\qn\ql\qm;\qn'\ql'\qm'}^{\mathfrak{l}\mathfrak{m}}(t;s)
    \left[
    \dif V_{\qn\ql\qm;\qn'\ql'\qm'}^{\mathfrak{l}\mathfrak{m}}(t-\tau;s)
    \right]^*
    \right\rangle
    \\
    &=
    G^2m_\psi^2\MPBH\,
    \mathcal{K}_{\qn\ql;\qn'\ql'}^{\mathfrak{l}}(s)
    \dif s\,
    \ee^{-|\tau|/\tau_c(s)} \ .
    \end{aligned}
    \label{eq:pbh-multipole-shell-correlator}
\end{equation}
Here the reduced multipole kernel is
\begin{equation}
    \mathcal{K}_{\qn\ql;\qn'\ql'}^{\mathfrak{l}}(s)
    =
    s^2
    \bar{\rho}_{\PBH}(s)
    \left[
    \frac{4\pi}{2\mathfrak{l}+1}
    \int_0^{r_{\mathrm{cut}}}
    \dif r\,
    u_{\qn\ql}(r)u_{\qn'\ql'}(r)
    \frac{[\min(r,s)]^{\mathfrak{l}}}{[\max(r,s)]^{\mathfrak{l}+1}}
    \right]^2
    W_{\ql\,\mathfrak{l}\,\ql'} \ ,
    \label{eq:pbh-multipole-mode-variance-kernel}
\end{equation}
with the angular weight given by
\begin{equation}
    W_{\ql\,\mathfrak{l}\,\ql'}
    =
    \frac{1}{2\ql'+1}
    \sum_{\qm\,\mathfrak{m}\,\qm'}
    \left|
    \int \dif\Omega\,
    Y_{\ql\qm}^*
    Y_{\mathfrak{l}\mathfrak{m}}
    Y_{\ql'\qm'}
    \right|^2
    =
    \frac{(2\ql+1)(2\mathfrak{l}+1)}{4\pi}
    \begin{pmatrix}
        \ql & \mathfrak{l} & \ql' \\
        0 & 0 & 0
    \end{pmatrix}^2 \ .
    \label{eq:pbh-gaunt-weight}
\end{equation}
The parenthesized two-row object is a Wigner $3j$ symbol.
The selection rules in Eq.~\eqref{eq:pbh-gaunt-weight} require $|\ql-\ql'|\le \mathfrak{l}\le\ql+\ql'$ and even $\ql+\mathfrak{l}+\ql'$.
For $\ql=\ql'=0$ and $\mathfrak{l}=0$, this kernel reduces to the monopole kernel $\mathcal{K}_{\qn\qn'}^{(0)}(s)$ in Eq.~\eqref{eq:pbh-monopole-mode-variance-kernel}.

Integrating over the shell contributions and Fourier transforming the exponential time dependence, we obtain the multipole contribution to the Hamiltonian matrix-element spectrum,
\begin{equation}
    S_{\qn\ql;\qn'\ql'}^{\mathfrak{l}}(\omega)
    =
    G^2m_\psi^2\MPBH
    \int_0^{r_{\mathrm{cut}}}
    \dif s\,
    \frac{
    2\tau_c(s)
    }{
    1+\omega^2\tau_c^2(s)
    }
    \mathcal{K}_{\qn\ql;\qn'\ql'}^{\mathfrak{l}}(s) \ .
    \label{eq:pbh-multipole-spectrum}
\end{equation}
The total spectrum for the transition between the two multiplets is then obtained by summing over the allowed PBH potential multipoles,
\begin{equation}
    S_{\qn\ql;\qn'\ql'}(\omega)
    =
    \sum_{\mathfrak{l}}
    S_{\qn\ql;\qn'\ql'}^{\mathfrak{l}}(\omega) \ ,
    \label{eq:pbh-multipole-total-spectrum}
\end{equation}
and the corresponding multiplet transition-rate estimate is
\begin{equation}
    \Gamma_{\qn'\ql'\to\qn\ql}
    \simeq
    \frac{1}{\hbar^2}
    S_{\qn\ql;\qn'\ql'}(\omega_{\qn\ql;\qn'\ql'}) \ ,
    \qquad
    \omega_{\qn\ql;\qn'\ql'}
    =
    \frac{E_{\qn\ql}-E_{\qn'\ql'}}{\hbar} \ .
    \label{eq:pbh-multipole-transition-rate}
\end{equation}

\begin{figure}[htbp]
    \centering
    \includegraphics{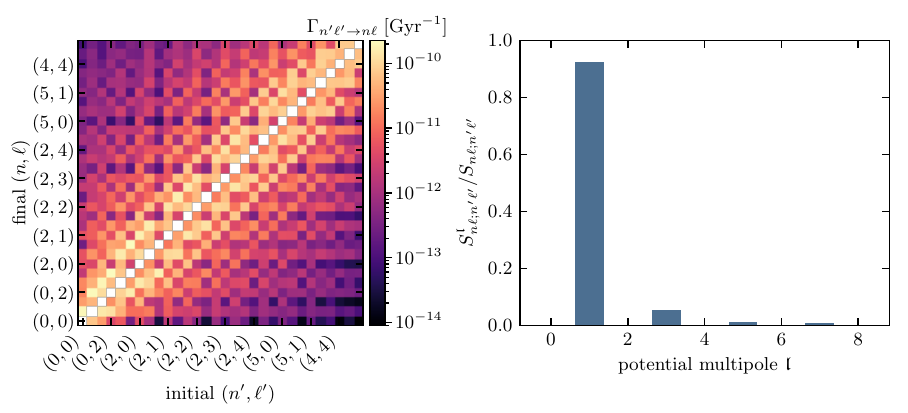}
    \caption{%
        Multiplet transition-rate estimate for the fiducial mixed halo with $\MPBH=1\Msun$.
        The plotted basis includes $\qn=0,\ldots,5$ for each $\ql=0,\ldots,4$, and the PBH potential expansion is truncated at $\mathfrak{l}=8$.
        Left: multiplet transition rates.
        Right: contribution of each PBH potential multipole $\mathfrak{l}$ to the transition spectrum for the fastest transition $(\qn',\ql')=(0,3)\to(\qn,\ql)=(0,4)$.
        This transition is driven mainly by the dipole potential fluctuation, and has a transition time $\simeq4.4\times10^9\,\mathrm{Gyr}$.
    }
    \label{fig:pbh-multipole-transition-rates}
\end{figure}

Figure~\ref{fig:pbh-multipole-transition-rates} shows that adding the angular multipoles increases the fastest low-lying-state rate by a factor of a few relative to the monopole benchmark.
The fastest transition time becomes of order $10^9\,\mathrm{Gyr}$ for $\MPBH=1\Msun$, while the median over off-diagonal transitions in the plotted basis is of order $10^{11}\,\mathrm{Gyr}$.
Since these rates scale linearly with $\MPBH$, the fastest plotted multiplet transition would require $\MPBH\sim10^8\Msun$ to become comparable to the age of the Universe.
As in the monopole case, this large-$\MPBH$ extrapolation should only be viewed as an illustrative extension of the perturbative estimate.
Thus the effect of PBH shot-noise fluctuations is enhanced by including the angular multipoles, but remains negligible for stellar-mass PBHs in the fiducial mixed halo.

Ref.~\cite{Boldrini:2019isx} studied PBH-induced cusp-to-core evolution in CDM halos and reported core-formation times of $1$--$8\,\mathrm{Gyr}$.
At first sight, these timescales appear much shorter than the ULDM transition and heating times obtained above.
However, that work considered initially cuspy $10^7\Msun$ halos containing $25$--$100\Msun$ PBHs, whereas our fiducial $10^{10}\Msun$ halo has a cored PBH profile and our calculation treats PBH shot noise perturbatively about a fixed background.
Applying their local core-formation estimate to our fiducial profile at $r=10\pc$ for $\MPBH=100\Msun$ gives the core-formation time $t_{\mathrm{c}}\simeq1.4\times10^8\,\mathrm{Gyr}$.
At the same PBH mass, our ULDM soliton heating time including angular multipoles is $t_{\mathrm{heat}}\simeq5.15\times10^8\,\mathrm{Gyr}$, which is comparable in order of magnitude to that estimate.
Thus, much of the apparent contrast can arise from the different halo regimes, although the two timescales characterize different structural responses.

\section{Conclusions}
\label{sec:conclusions}

Dark matter candidates span many orders of magnitude in mass, from ultralight bosonic fields to massive compact objects.
In this work, we connect these two extremes by investigating ULDM mixed with PBHs.
We developed a framework for dark matter halos containing a dominant ULDM component and a subdominant population of PBHs.
Starting from the coupled Schr\"odinger-Poisson and Newtonian equations, we separated the system into a time-independent averaged background and fluctuations around that background.
This formulation distinguishes two physically different PBH effects.
In the continuum limit, the PBH population contributes a background density to the gravitational potential.
For nonzero individual PBH mass, the same density profile is sampled by compact objects and produces a shot-noise fluctuation that can be projected onto the ULDM eigenmodes.

The background calculation is carried out for a spherical benchmark in which the ULDM density is modeled by a soliton--envelope profile and the PBH density by a cored-NFW profile.
For the fiducial mixed halo with $m_\psi=10^{-22}\eV$, $M_{\mathrm{vir}}\simeq10^{10}\Msun$, and $\fPBH=0.01$, the continuum PBH component changes the background density and gravitational potential only at the sub-percent level, and the low-lying eigenvalue shifts remain below about $0.25\%$.
The fitted eigenmode expansion nevertheless shows a more pronounced response in the coherent inner mode content.
The target ULDM density is reconstructed with a fractional mismatch $\sqrt{D}\simeq9.1\times10^{-4}$ over the fitted range, and the resulting random-phase ensemble shows that the continuum PBH contribution changes the radial participation measure in the coherent soliton region by about $20\%$.
The outer halo, by contrast, remains close to the many-mode random-wave regime.
Thus a percent-level PBH abundance has little effect on the spherically averaged halo profile, but it can alter the ULDM mode composition that builds the soliton region.

For the shot-noise fluctuations generated by the discrete PBH population, the Poisson estimate gives a coarse-grained density fluctuation that scales as $\MPBH^{1/2}$ at fixed background PBH density, and the associated monopole potential fluctuation is of order $10^{-6}$ of the background potential for $\MPBH=1\Msun$ in the fiducial halo.
Projecting this potential fluctuation onto the ULDM eigenbasis gives transition rates that are far below any relevant galactic rate.
For a smoothing length $\lambda=1\kpc$, the fastest monopole transition among the first eleven $\ql=0$ modes has a timescale of order $10^{10}\,\mathrm{Gyr}$, while the corresponding soliton-region heating estimate gives $t_{\mathrm{heat}}^{(0)}$ of order $10^{11}\,\mathrm{Gyr}$.
Including angular multipoles strengthens the rate estimate but does not change the physical conclusion.
In the low-lying multipole basis considered here, the fastest channel is $(\qn',\ql')=(0,3)\to(\qn,\ql)=(0,4)$ and is driven mainly by the dipole component of the PBH potential fluctuation.
Its transition time is of order $10^9\,\mathrm{Gyr}$ for $\fPBH=0.01$ and $\MPBH=1\Msun$, with a median off-diagonal transition time of order $10^{11}\,\mathrm{Gyr}$ in the plotted basis.
For stellar-mass PBHs in the fiducial halo, mode mixing driven by PBH shot-noise fluctuations is therefore negligible.
Because the rates are set by the PBH shot-noise variance, they scale linearly with $\MPBH$ at fixed PBH abundance, or equivalently with $\fPBH\MPBH$ in the perturbative regime.

The approximations used in the discrete-PBH calculation define the domain of this result.
The PBHs are treated as bare compact perturbers, without a correlated ULDM dressing, and their unequal-time correlations are modeled locally with an exponential decay set by the isotropic Jeans velocity dispersion.
The transition rates are also computed in a fixed background eigenbasis, rather than in a fully nonlinear, self-consistent time evolution.
An orbit-based PBH density correlator and direct Schr\"odinger-Poisson simulations in boosted-discreteness regimes would provide useful tests and extensions of the present calculation.
These refinements are important for precision modeling, but the enormous separation between the fiducial transition times and galactic timescales makes the suppression of stellar-mass PBH discreteness a robust outcome of the benchmark studied here.

In summary, this work provides a practical framework for treating mixed ULDM--PBH halos in terms of two distinct regimes.
The continuum structural response is controlled by the PBH abundance $\fPBH$, while the discrete PBH shot-noise effect is controlled by the product $\fPBH\MPBH$.
For the fiducial subdominant PBH abundance and stellar-mass PBHs, PBHs primarily act as a continuum in the ULDM halo, while their discreteness produces an extremely slow perturbative correction to ULDM mode evolution.

\section*{Acknowledgments}
X.-Y.~Y. is supported in part by the KIAS Individual Grant No.~QP090702.

\bibliographystyle{JHEP}
\bibliography{citeLib}

@article{Planck:2018vyg,
    author = "Aghanim, N. and others",
    collaboration = "Planck",
    title = "{Planck 2018 results. VI. Cosmological parameters}",
    eprint = "1807.06209",
    archivePrefix = "arXiv",
    primaryClass = "astro-ph.CO",
    doi = "10.1051/0004-6361/201833910",
    journal = "Astron. Astrophys.",
    volume = "641",
    pages = "A6",
    year = "2020",
    note = "[Erratum: Astron.Astrophys. 652, C4 (2021)]"
}

@article{Zwicky:1937zza,
    author = "Zwicky, F.",
    title = "{On the Masses of Nebulae and of Clusters of Nebulae}",
    doi = "10.1086/143864",
    journal = "Astrophys. J.",
    volume = "86",
    pages = "217--246",
    year = "1937"
}

@article{Rubin:1970zza,
    author = "Rubin, Vera C. and Ford, Jr., W. Kent",
    title = "{Rotation of the Andromeda Nebula from a Spectroscopic Survey of Emission Regions}",
    doi = "10.1086/150317",
    journal = "Astrophys. J.",
    volume = "159",
    pages = "379--403",
    year = "1970"
}

@article{Clowe:2006eq,
    author = "Clowe, Douglas and Bradac, Marusa and Gonzalez, Anthony H. and Markevitch, Maxim and Randall, Scott W. and Jones, Christine and Zaritsky, Dennis",
    title = "{A direct empirical proof of the existence of dark matter}",
    eprint = "astro-ph/0608407",
    archivePrefix = "arXiv",
    reportNumber = "SLAC-PUB-12078",
    doi = "10.1086/508162",
    journal = "Astrophys. J. Lett.",
    volume = "648",
    pages = "L109--L113",
    year = "2006"
}

@article{Bertone:2004pz,
    author = "Bertone, Gianfranco and Hooper, Dan and Silk, Joseph",
    title = "{Particle dark matter: Evidence, candidates and constraints}",
    eprint = "hep-ph/0404175",
    archivePrefix = "arXiv",
    reportNumber = "FERMILAB-PUB-04-047-A",
    doi = "10.1016/j.physrep.2004.08.031",
    journal = "Phys. Rept.",
    volume = "405",
    pages = "279--390",
    year = "2005"
}

@article{Feng:2010gw,
    author = "Feng, Jonathan L.",
    title = "{Dark Matter Candidates from Particle Physics and Methods of Detection}",
    eprint = "1003.0904",
    archivePrefix = "arXiv",
    primaryClass = "astro-ph.CO",
    reportNumber = "UCI-TR-2009-13",
    doi = "10.1146/annurev-astro-082708-101659",
    journal = "Ann. Rev. Astron. Astrophys.",
    volume = "48",
    pages = "495--545",
    year = "2010"
}

@article{Cirelli:2024ssz,
    author = "Cirelli, Marco and Strumia, Alessandro and Zupan, Jure",
    title = "{Dark Matter}",
    eprint = "2406.01705",
    archivePrefix = "arXiv",
    primaryClass = "hep-ph",
    doi = "10.21468/SciPostPhysRev.1",
    month = "6",
    year = "2024"
}

@article{Zurek:2008qg,
    author = "Zurek, Kathryn M.",
    title = "{Multi-Component Dark Matter}",
    eprint = "0811.4429",
    archivePrefix = "arXiv",
    primaryClass = "hep-ph",
    reportNumber = "FERMILAB-PUB-08-542-A",
    doi = "10.1103/PhysRevD.79.115002",
    journal = "Phys. Rev. D",
    volume = "79",
    pages = "115002",
    year = "2009"
}

@article{Kim:2023onk,
    author = "Kim, Jeong Han and Kong, Kyoungchul and Lim, Se Hwan and Park, Jong-Chul",
    title = "{Warm Surprises from Cold Duets: N-Body Simulations with Two-Component Dark Matter}",
    eprint = "2312.07660",
    archivePrefix = "arXiv",
    primaryClass = "hep-ph",
    doi = "10.1093/ptep/ptae169",
    journal = "PTEP",
    volume = "2024",
    number = "12",
    pages = "123B03",
    year = "2024"
}

@article{Marsh:2013ywa,
    author = "Marsh, David J. E. and Silk, Joe",
    title = "{A Model For Halo Formation With Axion Mixed Dark Matter}",
    eprint = "1307.1705",
    archivePrefix = "arXiv",
    primaryClass = "astro-ph.CO",
    doi = "10.1093/mnras/stt2079",
    journal = "Mon. Not. Roy. Astron. Soc.",
    volume = "437",
    number = "3",
    pages = "2652--2663",
    year = "2014"
}

@article{Cai:2020fnq,
    author = "Cai, Rong-Gen and Ding, Yu-Chen and Yang, Xing-Yu and Zhou, Yu-Feng",
    title = "{Constraints on a mixed model of dark matter particles and primordial black holes from the galactic 511 keV line}",
    eprint = "2007.11804",
    archivePrefix = "arXiv",
    primaryClass = "astro-ph.CO",
    doi = "10.1088/1475-7516/2021/03/057",
    journal = "JCAP",
    volume = "03",
    pages = "057",
    year = "2021"
}

@article{Cai:2022kbp,
    author = "Cai, Rong-Gen and Chen, Tan and Wang, Shao-Jiang and Yang, Xing-Yu",
    title = "{Gravitational microlensing by dressed primordial black holes}",
    eprint = "2210.02078",
    archivePrefix = "arXiv",
    primaryClass = "astro-ph.CO",
    doi = "10.1088/1475-7516/2023/03/043",
    journal = "JCAP",
    volume = "03",
    pages = "043",
    year = "2023"
}

@article{Preskill:1982cy,
    author = "Preskill, John and Wise, Mark B. and Wilczek, Frank",
    editor = "Srednicki, M. A.",
    title = "{Cosmology of the Invisible Axion}",
    reportNumber = "HUTP-82-A048, NSF-ITP-82-103",
    doi = "10.1016/0370-2693(83)90637-8",
    journal = "Phys. Lett. B",
    volume = "120",
    pages = "127--132",
    year = "1983"
}

@article{Abbott:1982af,
    author = "Abbott, L. F. and Sikivie, P.",
    editor = "Srednicki, M. A.",
    title = "{A Cosmological Bound on the Invisible Axion}",
    reportNumber = "PRINT-82-0695 (BRANDEIS)",
    doi = "10.1016/0370-2693(83)90638-X",
    journal = "Phys. Lett. B",
    volume = "120",
    pages = "133--136",
    year = "1983"
}

@article{Hu:2000ke,
    author = "Hu, Wayne and Barkana, Rennan and Gruzinov, Andrei",
    title = "{Cold and fuzzy dark matter}",
    eprint = "astro-ph/0003365",
    archivePrefix = "arXiv",
    doi = "10.1103/PhysRevLett.85.1158",
    journal = "Phys. Rev. Lett.",
    volume = "85",
    pages = "1158--1161",
    year = "2000"
}

@article{Hui:2016ltb,
    author = "Hui, Lam and Ostriker, Jeremiah P. and Tremaine, Scott and Witten, Edward",
    title = "{Ultralight scalars as cosmological dark matter}",
    eprint = "1610.08297",
    archivePrefix = "arXiv",
    primaryClass = "astro-ph.CO",
    doi = "10.1103/PhysRevD.95.043541",
    journal = "Phys. Rev. D",
    volume = "95",
    number = "4",
    pages = "043541",
    year = "2017"
}

@article{Ferreira:2020fam,
    author = "Ferreira, Elisa G. M.",
    title = "{Ultra-light dark matter}",
    eprint = "2005.03254",
    archivePrefix = "arXiv",
    primaryClass = "astro-ph.CO",
    doi = "10.1007/s00159-021-00135-6",
    journal = "Astron. Astrophys. Rev.",
    volume = "29",
    number = "1",
    pages = "7",
    year = "2021"
}

@article{Hui:2021tkt,
    author = "Hui, Lam",
    title = "{Wave Dark Matter}",
    eprint = "2101.11735",
    archivePrefix = "arXiv",
    primaryClass = "astro-ph.CO",
    doi = "10.1146/annurev-astro-120920-010024",
    journal = "Ann. Rev. Astron. Astrophys.",
    volume = "59",
    pages = "247--289",
    year = "2021"
}

@article{Widrow:1993qq,
    author = "Widrow, Lawrence M. and Kaiser, Nick",
    title = "{Using the Schrodinger equation to simulate collisionless matter}",
    reportNumber = "CITA-93-17",
    journal = "Astrophys. J. Lett.",
    volume = "416",
    pages = "L71--L74",
    year = "1993"
}

@article{Navarro:1996gj,
    author = "Navarro, Julio F. and Frenk, Carlos S. and White, Simon D. M.",
    title = "{A Universal density profile from hierarchical clustering}",
    eprint = "astro-ph/9611107",
    archivePrefix = "arXiv",
    doi = "10.1086/304888",
    journal = "Astrophys. J.",
    volume = "490",
    pages = "493--508",
    year = "1997"
}

@article{Schive:2014dra,
    author = "Schive, Hsi-Yu and Chiueh, Tzihong and Broadhurst, Tom",
    title = "{Cosmic Structure as the Quantum Interference of a Coherent Dark Wave}",
    eprint = "1406.6586",
    archivePrefix = "arXiv",
    primaryClass = "astro-ph.GA",
    doi = "10.1038/nphys2996",
    journal = "Nature Phys.",
    volume = "10",
    pages = "496--499",
    year = "2014"
}

@article{Schive:2014hza,
    author = "Schive, Hsi-Yu and Liao, Ming-Hsuan and Woo, Tak-Pong and Wong, Shing-Kwong and Chiueh, Tzihong and Broadhurst, Tom and Hwang, W. -Y. Pauchy",
    title = "{Understanding the Core-Halo Relation of Quantum Wave Dark Matter from 3D Simulations}",
    eprint = "1407.7762",
    archivePrefix = "arXiv",
    primaryClass = "astro-ph.GA",
    doi = "10.1103/PhysRevLett.113.261302",
    journal = "Phys. Rev. Lett.",
    volume = "113",
    number = "26",
    pages = "261302",
    year = "2014"
}

@article{Hawking:1971ei,
    author = "Hawking, Stephen",
    title = "{Gravitationally collapsed objects of very low mass}",
    doi = "10.1093/mnras/152.1.75",
    journal = "Mon. Not. Roy. Astron. Soc.",
    volume = "152",
    pages = "75",
    year = "1971"
}

@article{Carr:1974nx,
    author = "Carr, Bernard J. and Hawking, S. W.",
    title = "{Black holes in the early Universe}",
    doi = "10.1093/mnras/168.2.399",
    journal = "Mon. Not. Roy. Astron. Soc.",
    volume = "168",
    pages = "399--415",
    year = "1974"
}

@article{Afshordi:2003zb,
    author = "Afshordi, N. and McDonald, P. and Spergel, D. N.",
    title = "{Primordial black holes as dark matter: The Power spectrum and evaporation of early structures}",
    eprint = "astro-ph/0302035",
    archivePrefix = "arXiv",
    doi = "10.1086/378763",
    journal = "Astrophys. J. Lett.",
    volume = "594",
    pages = "L71--L74",
    year = "2003"
}

@article{LIGOScientific:2016aoc,
    author = "Abbott, B. P. and others",
    collaboration = "LIGO Scientific, Virgo",
    title = "{Observation of Gravitational Waves from a Binary Black Hole Merger}",
    eprint = "1602.03837",
    archivePrefix = "arXiv",
    primaryClass = "gr-qc",
    reportNumber = "LIGO-P150914",
    doi = "10.1103/PhysRevLett.116.061102",
    journal = "Phys. Rev. Lett.",
    volume = "116",
    number = "6",
    pages = "061102",
    year = "2016"
}

@article{Bird:2016dcv,
    author = {Bird, Simeon and Cholis, Ilias and Mu{\~n}oz, Julian B. and Ali-Ha{\"\i}moud, Yacine and Kamionkowski, Marc and Kovetz, Ely D. and Raccanelli, Alvise and Riess, Adam G.},
    title = "{Did LIGO detect dark matter?}",
    eprint = "1603.00464",
    archivePrefix = "arXiv",
    primaryClass = "astro-ph.CO",
    doi = "10.1103/PhysRevLett.116.201301",
    journal = "Phys. Rev. Lett.",
    volume = "116",
    number = "20",
    pages = "201301",
    year = "2016"
}

@article{Clesse:2016vqa,
    author = "Clesse, Sebastien and Garc{\'\i}a-Bellido, Juan",
    title = "{The clustering of massive Primordial Black Holes as Dark Matter: measuring their mass distribution with Advanced LIGO}",
    eprint = "1603.05234",
    archivePrefix = "arXiv",
    primaryClass = "astro-ph.CO",
    reportNumber = "TTK-16-10, IFT-UAM-CSIC-16-027",
    doi = "10.1016/j.dark.2016.10.002",
    journal = "Phys. Dark Univ.",
    volume = "15",
    pages = "142--147",
    year = "2017"
}

@article{Sasaki:2016jop,
    author = "Sasaki, Misao and Suyama, Teruaki and Tanaka, Takahiro and Yokoyama, Shuichiro",
    title = "{Primordial Black Hole Scenario for the Gravitational-Wave Event GW150914}",
    eprint = "1603.08338",
    archivePrefix = "arXiv",
    primaryClass = "astro-ph.CO",
    reportNumber = "RESCEU-17-16, RUP-16-7, YITP-16-43",
    doi = "10.1103/PhysRevLett.117.061101",
    journal = "Phys. Rev. Lett.",
    volume = "117",
    number = "6",
    pages = "061101",
    year = "2016",
    note = "[Erratum: Phys.Rev.Lett. 121, 059901 (2018)]"
}

@article{Carr:2016drx,
    author = "Carr, Bernard and Kuhnel, Florian and Sandstad, Marit",
    title = "{Primordial Black Holes as Dark Matter}",
    eprint = "1607.06077",
    archivePrefix = "arXiv",
    primaryClass = "astro-ph.CO",
    reportNumber = "NORDITA-2016-83",
    doi = "10.1103/PhysRevD.94.083504",
    journal = "Phys. Rev. D",
    volume = "94",
    number = "8",
    pages = "083504",
    year = "2016"
}

@article{Ali-Haimoud:2016mbv,
    author = {Ali-Ha{\"\i}moud, Yacine and Kamionkowski, Marc},
    title = "{Cosmic microwave background limits on accreting primordial black holes}",
    eprint = "1612.05644",
    archivePrefix = "arXiv",
    primaryClass = "astro-ph.CO",
    doi = "10.1103/PhysRevD.95.043534",
    journal = "Phys. Rev. D",
    volume = "95",
    number = "4",
    pages = "043534",
    year = "2017"
}

@article{Niikura:2017zjd,
    author = "Niikura, Hiroko and others",
    title = "{Microlensing constraints on primordial black holes with Subaru/HSC Andromeda observations}",
    eprint = "1701.02151",
    archivePrefix = "arXiv",
    primaryClass = "astro-ph.CO",
    doi = "10.1038/s41550-019-0723-1",
    journal = "Nature Astron.",
    volume = "3",
    number = "6",
    pages = "524--534",
    year = "2019"
}

@article{Ali-Haimoud:2017rtz,
    author = {Ali-Ha{\"\i}moud, Yacine and Kovetz, Ely D. and Kamionkowski, Marc},
    title = "{Merger rate of primordial black-hole binaries}",
    eprint = "1709.06576",
    archivePrefix = "arXiv",
    primaryClass = "astro-ph.CO",
    doi = "10.1103/PhysRevD.96.123523",
    journal = "Phys. Rev. D",
    volume = "96",
    number = "12",
    pages = "123523",
    year = "2017"
}

@article{Sasaki:2018dmp,
    author = "Sasaki, Misao and Suyama, Teruaki and Tanaka, Takahiro and Yokoyama, Shuichiro",
    title = "{Primordial black holes{\textemdash}perspectives in gravitational wave astronomy}",
    eprint = "1801.05235",
    archivePrefix = "arXiv",
    primaryClass = "astro-ph.CO",
    doi = "10.1088/1361-6382/aaa7b4",
    journal = "Class. Quant. Grav.",
    volume = "35",
    number = "6",
    pages = "063001",
    year = "2018"
}

@article{Cai:2019elf,
    author = "Cai, Rong-Gen and Pi, Shi and Wang, Shao-Jiang and Yang, Xing-Yu",
    title = "{Pulsar Timing Array Constraints on the Induced Gravitational Waves}",
    eprint = "1907.06372",
    archivePrefix = "arXiv",
    primaryClass = "astro-ph.CO",
    doi = "10.1088/1475-7516/2019/10/059",
    journal = "JCAP",
    volume = "10",
    pages = "059",
    year = "2019"
}

@article{Chen:2019xse,
    author = "Chen, Zu-Cheng and Yuan, Chen and Huang, Qing-Guo",
    title = "{Pulsar Timing Array Constraints on Primordial Black Holes with NANOGrav 11-Year Dataset}",
    eprint = "1910.12239",
    archivePrefix = "arXiv",
    primaryClass = "astro-ph.CO",
    doi = "10.1103/PhysRevLett.124.251101",
    journal = "Phys. Rev. Lett.",
    volume = "124",
    number = "25",
    pages = "25",
    year = "2020"
}

@article{Cai:2019bmk,
    author = "Cai, Rong-Gen and Guo, Zong-Kuan and Liu, Jing and Liu, Lang and Yang, Xing-Yu",
    title = "{Primordial black holes and gravitational waves from parametric amplification of curvature perturbations}",
    eprint = "1912.10437",
    archivePrefix = "arXiv",
    primaryClass = "astro-ph.CO",
    doi = "10.1088/1475-7516/2020/06/013",
    journal = "JCAP",
    volume = "06",
    pages = "013",
    year = "2020"
}

@article{Carr:2020gox,
    author = "Carr, Bernard and Kohri, Kazunori and Sendouda, Yuuiti and Yokoyama, Jun'ichi",
    title = "{Constraints on primordial black holes}",
    eprint = "2002.12778",
    archivePrefix = "arXiv",
    primaryClass = "astro-ph.CO",
    reportNumber = "RESCEU-03/20; KEK-Cosmo-249; KEK-TH-2199; IPMU20-0024",
    doi = "10.1088/1361-6633/ac1e31",
    journal = "Rept. Prog. Phys.",
    volume = "84",
    number = "11",
    pages = "116902",
    year = "2021"
}

@article{Carr:2020xqk,
    author = "Carr, Bernard and Kuhnel, Florian",
    title = "{Primordial Black Holes as Dark Matter: Recent Developments}",
    eprint = "2006.02838",
    archivePrefix = "arXiv",
    primaryClass = "astro-ph.CO",
    doi = "10.1146/annurev-nucl-050520-125911",
    journal = "Ann. Rev. Nucl. Part. Sci.",
    volume = "70",
    pages = "355--394",
    year = "2020"
}

@article{Green:2020jor,
    author = "Green, Anne M. and Kavanagh, Bradley J.",
    title = "{Primordial Black Holes as a dark matter candidate}",
    eprint = "2007.10722",
    archivePrefix = "arXiv",
    primaryClass = "astro-ph.CO",
    doi = "10.1088/1361-6471/abc534",
    journal = "J. Phys. G",
    volume = "48",
    number = "4",
    pages = "043001",
    year = "2021"
}

@article{DeLuca:2020agl,
    author = "De Luca, V. and Franciolini, G. and Riotto, A.",
    title = "{NANOGrav Data Hints at Primordial Black Holes as Dark Matter}",
    eprint = "2009.08268",
    archivePrefix = "arXiv",
    primaryClass = "astro-ph.CO",
    doi = "10.1103/PhysRevLett.126.041303",
    journal = "Phys. Rev. Lett.",
    volume = "126",
    number = "4",
    pages = "041303",
    year = "2021"
}

@article{Kohri:2020qqd,
    author = "Kohri, Kazunori and Terada, Takahiro",
    title = "{Solar-Mass Primordial Black Holes Explain NANOGrav Hint of Gravitational Waves}",
    eprint = "2009.11853",
    archivePrefix = "arXiv",
    primaryClass = "astro-ph.CO",
    reportNumber = "KEK-TH-2260, KEK-Cosmo-0263, CTPU-PTC-20-22",
    doi = "10.1016/j.physletb.2020.136040",
    journal = "Phys. Lett. B",
    volume = "813",
    pages = "136040",
    year = "2021"
}

@article{Yuan:2021qgz,
    author = "Yuan, Chen and Huang, Qing-Guo",
    title = "{A topic review on probing primordial black hole dark matter with scalar induced gravitational waves}",
    eprint = "2103.04739",
    archivePrefix = "arXiv",
    primaryClass = "astro-ph.GA",
    doi = "10.1016/j.isci.2021.102860",
    journal = "iScience",
    volume = "24",
    pages = "102860",
    year = "2021"
}

@article{Liu:2021jnw,
    author = "Liu, Lang and Yang, Xing-Yu and Guo, Zong-Kuan and Cai, Rong-Gen",
    title = "{Testing primordial black hole and measuring the Hubble constant with multiband gravitational-wave observations}",
    eprint = "2112.05473",
    archivePrefix = "arXiv",
    primaryClass = "astro-ph.CO",
    doi = "10.1088/1475-7516/2023/01/006",
    journal = "JCAP",
    volume = "01",
    pages = "006",
    year = "2023"
}

@article{Escriva:2022duf,
    author = "Escriv{\`a}, Albert and Kuhnel, Florian and Tada, Yuichiro",
    editor = "Sedda, Manuel Arca and Bortolas, Elisa and Spera, Mario",
    title = "{Primordial Black Holes}",
    eprint = "2211.05767",
    archivePrefix = "arXiv",
    primaryClass = "astro-ph.CO",
    doi = "10.1016/B978-0-32-395636-9.00012-8",
    month = "11",
    year = "2022"
}

@article{Ozsoy:2023ryl,
    author = {{\"O}zsoy, Ogan and Tasinato, Gianmassimo},
    title = "{Inflation and Primordial Black Holes}",
    eprint = "2301.03600",
    archivePrefix = "arXiv",
    primaryClass = "astro-ph.CO",
    doi = "10.3390/universe9050203",
    journal = "Universe",
    volume = "9",
    number = "5",
    pages = "203",
    year = "2023"
}

@article{LISACosmologyWorkingGroup:2023njw,
    author = "Bagui, Eleni and others",
    collaboration = "LISA Cosmology Working Group",
    title = "{Primordial black holes and their gravitational-wave signatures}",
    eprint = "2310.19857",
    archivePrefix = "arXiv",
    primaryClass = "astro-ph.CO",
    doi = "10.1007/s41114-024-00053-w",
    journal = "Living Rev. Rel.",
    volume = "28",
    number = "1",
    pages = "1",
    year = "2025"
}

@article{Yavetz:2021pbc,
    author = "Yavetz, Tomer D. and Li, Xinyu and Hui, Lam",
    title = "{Construction of wave dark matter halos: Numerical algorithm and analytical constraints}",
    eprint = "2109.06125",
    archivePrefix = "arXiv",
    primaryClass = "astro-ph.CO",
    doi = "10.1103/PhysRevD.105.023512",
    journal = "Phys. Rev. D",
    volume = "105",
    number = "2",
    pages = "023512",
    year = "2022"
}

@article{Schwabe:2020eac,
    author = "Schwabe, Bodo and Gosenca, Mateja and Behrens, Christoph and Niemeyer, Jens C. and Easther, Richard",
    title = "{Simulating mixed fuzzy and cold dark matter}",
    eprint = "2007.08256",
    archivePrefix = "arXiv",
    primaryClass = "astro-ph.CO",
    doi = "10.1103/PhysRevD.102.083518",
    journal = "Phys. Rev. D",
    volume = "102",
    number = "8",
    pages = "083518",
    year = "2020"
}

@article{Afshordi:2026arr,
    author = "Afshordi, Niayesh and Halper, Phil and Rini, Matteo and Schirber, Michael",
    title = "{Big Mysteries Survey: Physicists' Views on Cosmology, Black Holes, Quantum Mechanics, and Quantum Gravity}",
    eprint = "2605.11058",
    archivePrefix = "arXiv",
    primaryClass = "physics.soc-ph",
    month = "5",
    year = "2026"
}

@BOOK{2008gady.book.....B,
    author = {{Binney}, James and {Tremaine}, Scott},
    title = "{Galactic Dynamics: Second Edition}",
    year = 2008,
    publisher = {Princeton University Press},
    adsurl = {https://ui.adsabs.harvard.edu/abs/2008gady.book.....B}
}

@article{Dong:2015yjs,
    author = "Dong, Ruifeng and Kinney, William H and Stojkovic, Dejan",
    title = "{Gravitational wave production by Hawking radiation from rotating primordial black holes}",
    eprint = "1511.05642",
    archivePrefix = "arXiv",
    primaryClass = "astro-ph.CO",
    doi = "10.1088/1475-7516/2016/10/034",
    journal = "JCAP",
    volume = "10",
    pages = "034",
    year = "2016"
}

@article{Khlopov:1985fch,
    author = "Khlopov, M. Yu. and Malomed, B. A. and Zeldovich, Ia. B. and Zeldovich, Ya. B.",
    title = "{Gravitational instability of scalar fields and formation of primordial black holes}",
    doi = "10.1093/mnras/215.4.575",
    journal = "Mon. Not. Roy. Astron. Soc.",
    volume = "215",
    number = "4",
    pages = "575--589",
    year = "1985"
}

@article{Khlopov:2008qy,
    author = "Khlopov, Maxim Yu.",
    title = "{Primordial Black Holes}",
    eprint = "0801.0116",
    archivePrefix = "arXiv",
    primaryClass = "astro-ph",
    doi = "10.1088/1674-4527/10/6/001",
    journal = "Res. Astron. Astrophys.",
    volume = "10",
    pages = "495--528",
    year = "2010"
}

@article{Belotsky:2018wph,
    author = "Belotsky, Konstantin M. and Dokuchaev, Vyacheslav I. and Eroshenko, Yury N. and Esipova, Ekaterina A. and Khlopov, Maxim Yu. and Khromykh, Leonid A. and Kirillov, Alexander A. and Nikulin, Valeriy V. and Rubin, Sergey G. and Svadkovsky, Igor V.",
    title = "{Clusters of primordial black holes}",
    eprint = "1807.06590",
    archivePrefix = "arXiv",
    primaryClass = "astro-ph.CO",
    doi = "10.1140/epjc/s10052-019-6741-4",
    journal = "Eur. Phys. J. C",
    volume = "79",
    number = "3",
    pages = "246",
    year = "2019"
}

@article{Davoudiasl:2019nlo,
    author = "Davoudiasl, Hooman and Denton, Peter B",
    title = "{Ultralight Boson Dark Matter and Event Horizon Telescope Observations of M87*}",
    eprint = "1904.09242",
    archivePrefix = "arXiv",
    primaryClass = "astro-ph.CO",
    doi = "10.1103/PhysRevLett.123.021102",
    journal = "Phys. Rev. Lett.",
    volume = "123",
    number = "2",
    pages = "021102",
    year = "2019"
}

@article{Kimura:2021sqz,
    author = "Kimura, Rampei and Suyama, Teruaki and Yamaguchi, Masahide and Zhang, Ying-Li",
    title = "{Reconstruction of Primordial Power Spectrum of curvature perturbation from the merger rate of Primordial Black Hole Binaries}",
    eprint = "2102.05280",
    archivePrefix = "arXiv",
    primaryClass = "astro-ph.CO",
    doi = "10.1088/1475-7516/2021/04/031",
    journal = "JCAP",
    volume = "04",
    pages = "031",
    year = "2021"
}

@article{Wang:2022nml,
    author = "Wang, Xinpeng and Zhang, Ying-li and Kimura, Rampei and Yamaguchi, Masahide",
    title = "{Reconstruction of power spectrum of primordial curvature perturbations on small scales from primordial black hole binaries scenario of LIGO/VIRGO detection}",
    eprint = "2209.12911",
    archivePrefix = "arXiv",
    primaryClass = "astro-ph.CO",
    doi = "10.1007/s11433-023-2091-x",
    journal = "Sci. China Phys. Mech. Astron.",
    volume = "66",
    number = "6",
    pages = "260462",
    year = "2023"
}

@article{Yin:2024xov,
    author = "Yin, Ziwen and Visinelli, Luca",
    title = "{Axion star condensation around primordial black holes and microlensing limits}",
    eprint = "2404.10340",
    archivePrefix = "arXiv",
    primaryClass = "hep-ph",
    reportNumber = "CA21106; CA21136",
    doi = "10.1088/1475-7516/2024/10/013",
    journal = "JCAP",
    volume = "10",
    pages = "013",
    year = "2024"
}

@article{Hertzberg:2020hsz,
    author = "Hertzberg, Mark P. and Schiappacasse, Enrico D. and Yanagida, Tsutomu T.",
    title = "{Axion Star Nucleation in Dark Minihalos around Primordial Black Holes}",
    eprint = "2001.07476",
    archivePrefix = "arXiv",
    primaryClass = "astro-ph.CO",
    doi = "10.1103/PhysRevD.102.023013",
    journal = "Phys. Rev. D",
    volume = "102",
    number = "2",
    pages = "023013",
    year = "2020"
}

@article{Boldrini:2019isx,
    author = "Boldrini, Pierre and Miki, Yohei and Wagner, Alexander Y. and Mohayaee, Roya and Silk, Joseph and Arbey, Alexandre",
    title = "{Cusp-to-core transition in low-mass dwarf galaxies induced by dynamical heating of cold dark matter by primordial black holes}",
    eprint = "1909.07395",
    archivePrefix = "arXiv",
    primaryClass = "astro-ph.CO",
    doi = "10.1093/mnras/staa150",
    journal = "Mon. Not. Roy. Astron. Soc.",
    volume = "492",
    number = "4",
    pages = "5218--5225",
    year = "2020"
}

\end{document}